\documentclass[twocolumn,api,jcp]{revtex4-2}
\pdfoutput=1
\usepackage[utf8]{inputenc}
\usepackage[T1]{fontenc}
\usepackage{amssymb}
\usepackage{amsmath}
\usepackage{graphics}
\usepackage{graphicx}
\usepackage{subcaption}
\usepackage{color}
\usepackage{calc}
\usepackage[english]{babel}
\usepackage{listings}
\usepackage{mhchem}
\usepackage{svg}
\usepackage{hyperref}
\usepackage{tikz}
\usepackage{setspace}
\usepackage{multirow}
\usepackage{array}
\usepackage{import}
\usepackage{float}
\usepackage{bm}
\usepackage{physics}
\usepackage{braket}
\usepackage[shortcuts]{extdash}

\usepackage[normalem]{ulem}

\renewcommand{\vec}[1]{\bm{#1}}
\newcommand{\diagram}[2][0.5]{\raisebox{0.5ex-#1\height}{\includegraphics{#2}}}

\usetikzlibrary{decorations.pathmorphing}
\usetikzlibrary{decorations.markings}
\usetikzlibrary{3d,calc}
\usetikzlibrary{shapes.geometric}
\usetikzlibrary{positioning}
\usetikzlibrary{chains,snakes,shapes}

\begin{document}

\title{
Investigating the basis set convergence of
diagrammatically decomposed coupled\-/cluster correlation energy contributions for the uniform electron gas
}

\author{Nikolaos Masios}
\author{Felix Hummel}
\author{Andreas Gr\"uneis}
\author{Andreas Irmler}
\email{andreas.irmler@tuwien.ac.at}
\affiliation{
  Institute for Theoretical Physics, TU Wien,\\
  Wiedner Hauptstraße 8--10/136, 1040 Vienna, Austria
}
\date{\today, DRAFT}

\begin{abstract}
We investigate the convergence of coupled\-/cluster correlation energies and
related quantities with respect to the employed basis set size for the uniform
electron gas to gain a better understanding of the basis set incompleteness
error.  To this end, coupled\-/cluster doubles (CCD) theory is applied to the
three dimensional uniform electron gas for a range of densities, basis set
sizes and electron numbers.  We present a detailed analysis of individual,
diagrammatically decomposed contributions to the amplitudes at the level of CCD
theory.  In particular, we show that only two terms from the amplitude
equations contribute to the asymptotic large\-/momentum behavior of the
transition structure factor, corresponding to the cusp region at short
interelectronic distances.  However, due to the coupling present in the
amplitude equations, all decomposed correlation energy contributions show the
same asymptotic convergence behavior to the complete basis set limit.  These
findings provide an additional rationale for the success of a recently proposed
correction to the basis set incompleteness error (BSIE) of coupled\-/cluster
theory.
Lastly, we examine the BSIE in the coupled\-/cluster doubles plus perturbative
triples [CCD(T)] method, as well as in the newly proposed coupled\-/cluster
doubles plus complete perturbative triples [CCD(cT)] method.
\end{abstract}

\maketitle
%

\section{Introduction}

Coupled\-/cluster (CC) theories are widely used in molecular quantum chemistry
and become increasingly popular for studying solid state systems.
CC theories approximate the true many\-/electron wave function in a systematically
improvable manner by employing an exponential ansatz with a series of higher order
particle\-/hole excitation operators.
While systems exhibiting strong electronic correlation effects
require high\-/order excitation operators, systems with strong single\-/reference
character can be well described using low order excitation operators~\cite{bartlett2007}.
In particular, at the truncation level of  single, double, and
perturbative triple particle\-/hole excitation operators, CCSD(T) theory predicts
atomization and reaction energies for a large number of molecules with an
accuracy of approximately 1~kcal/mol~\cite{bartlett2007}.
Although the computational cost of CCSD(T) theory
is significantly larger than that of the more widely used approximate density
functional theory calculations, recent methodological developments
enable the study of relatively complex systems, for instance,
molecules adsorbed on surfaces~\cite{Voloshina2011,kubas2016,Brandenburg2019,tsatsoulis2018,Sauer2019,Lau2021,Schaefer2021,Mullan2022,Shi2023}.
However, high accuracy compared to experiment can only be achieved
if the ansatz is fully converged with respect to all computational
parameters that model the true physical system.
These include the number of atoms used to model an infinitely large periodic crystal and the
truncation parameter of the basis set.
Any truncation of the employed one\-/electron basis set introduces
a basis set incompleteness error (BSIE) in CC and related theories.

This work aims at a detailed investigation of the BSIE in CCSD and CCSD(T)
methods using a plane wave basis set. For a better understanding of the
corresponding BSIE we employ the uniform electron gas (UEG) model Hamiltonian,
which includes a kinetic energy operator, an electronic Coulomb interaction and
a constant background potential to preserve charge neutrality. The UEG model
depends only on parameters that have a well defined physical interpretation:
(i) the electronic density, (ii) the number of electrons and the cell shape,
and (iii) a momentum cutoff defining the employed basis set. The electronic
density controls the relative importance of the kinetic energy operator
compared to the Coulomb interaction. In this manner, one can continuously
transform the system from a weakly correlated system at high densities to a
strongly correlated system at low densities.  The number of electrons and the
cell shape used to model the electron gas at a fixed density, allow for the
investigation of finite size effects~\cite{Shepherd2013,Masios2023}.  Due to
the translational symmetry of the UEG model Hamiltonian, the mean\-/field
orbital solutions correspond to plane waves characterized by a wave vector. The
momentum cutoff makes it possible to truncate the employed plane wave basis in
a systematic manner, making the UEG ideally suited to study BSIEs of electronic
structure theories~\cite{shepherd2012,Grueneis2013,Callahan2021}.

In this work we are mainly interested in the BSIEs introduced by large
cutoff momenta compared to the Fermi momentum of the UEG.
Although the Fermi sphere defines a complete plane wave basis set needed
for the representation of the mean\-/field ground state wave function,
the representation of correlated
wave functions requires a significantly larger basis set.
In particular, at the point where two electrons coalescence, the exact correlated
wave function exhibits a cusp~\cite{kato_1957,Pack1966,Morgan1993}.
As a consequence, a large number of one\-/electron orbitals is needed to
describe this feature with sufficient accuracy.
Apart from increasing the one\-/electron basis set,
it is also possible to account for the cusp in the wave function
and derived properties directly by adding basis functions explicitly depending
on two electronic coordinates.
A variety of techniques have been developed to accelerate the slow
convergence to the complete basis set limit including explicitly correlated
methods~\cite{kutzelnigg_1991,Tenno2007,Haettig2012,Kong2012,Gruneis2017}, transcorrelated
methods~\cite{Boys1969a,Boys1969b,Liao_transcorrelated,TENNO2000169,Cohen2019}, or basis\-/set
extrapolation techniques~\cite{Feller2013}. These methods are mainly
used for molecular calculations.

Recently, we proposed and investigated a finite basis set correction for
the CCSD method that is based on a diagrammatic
decomposition of the correlation energy~\cite{Irmler2021}.  We have shown that the finite basis
set error is dominated by two contributions to the CCSD correlation energies,
corresponding to the second\-/order energy and a term that we referred to as the
particle--particle ladder (PPL) term~\cite{Irmler2019}.  In Ref.~\cite{Irmler2021}
we have examined the accuracy
of different approximate corrections to the basis set incompleteness error in
the PPL term. Here, we present a
more detailed study of the various correlation energy contributions
that can be obtained from a diagrammatic decomposition of the CCSD correlation
energy.  We also investigate other related quantities as functions of the basis set
size, the electron number, and the electronic density.  This investigation allows us to
determine the next\-/to\-/leading\-/order contributions to the basis set error in the UEG.

Our diagrammatic decomposition approach is partly motivated by prominent
examples that sum up particular contributions in the perturbation
series to infinite order. Famous examples are the resummation of ring\-/type
contributions, as demonstrated in the random phase approximation
(RPA)~\cite{pines_collective_1952}, and ladder theory
(LT)~\cite{Hede1972,yasuhara_1972}, which contains PPL contributions.
CCSD contains all diagrams appearing in RPA \emph{and}
LT, as well as many further contributions beyond that.
This feature alone makes CCSD interesting as both, RPA and LT, are known to have
prominent failures.
The RPA lacks an accurate description of the short\-/range electronic
correlation. Already at medium densities the pair\-/correlation function
gets negative for vanishing interelectronic distances~\cite{Singwi1968}.
On the other hand, the RPA is known for providing an accurate description
in the long\-/range regime~\cite{gell-mann_correlation_1957}.
In contrast, short\-/range electron correlation can be properly described by the LT~\cite{Qian2006}.
Apart from that, in the particular case of systems with long ranged Coulomb interactions,
LT was found to be unsatisfactory~\cite{cioslowski_2005}.

We will present the theoretical framework in section~\ref{sec:theory} and
describe the computational details in section~\ref{sec:compDetails}.
Section~\ref{sec:ccdbsie} is devoted to the BSIE of the coupled cluster doubles
(CCD) method and related theories, whereas section~\ref{sec:tbsie} discusses
the BSIE for the studied methods including triple excitations.

\section{Theory}
\label{sec:theory}
\subsection{The uniform electron gas}

\begin{figure}[t]
  \includegraphics{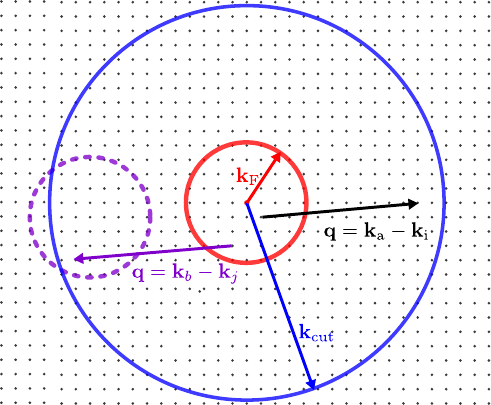}
\caption {Illustration of the reciprocal grid: Wave
vectors of occupied orbitals are found inside the Fermi sphere
of radius $k_\mathrm{F}$, shown as a red circle.
The wave vectors of virtual states in the finite basis are located outside the
Fermi sphere but inside the cutoff sphere of radius $k_\mathrm{cut}$,
shown as a blue circle.
The wave vectors of the remaining infinite augmented virtual
states are located beyond the blue sphere.
The black arrow indicates a one\-/body
excitation process with momentum transfer $\vec{q}$ from an occupied orbital
$\vec{k}_i$ to a virtual orbital $\vec{k}_a$.
The purple circle depicts the set
of all non\-/vanishing ${\vec k}_b$ in the four\-/index Coulomb integral $\upsilon_{ij}^{ab}$
for two given states $i$ and $a$ with the momentum transfer $\vec{q}=\vec{k}_a-\vec{k}_i$.
Note that depending on the momentum transfer $\vec{q}$, the set of non\-/vanishing
virtual states can be located either entirely inside, entirely outside, or
partially outside the radius $k_\mathrm{cut}$.}
\label{fig:rez-mesh}
\end{figure}


In this work, we study the UEG model system with $N$ electrons within a finite
simulation cell under periodic boundary conditions. A positive background
charge ensures charge neutrality in the unit cell.
We work with spatial orbitals, each occupied by two electrons with opposite spin. We will
restrict the analysis to cubic boxes, with a box volume $\Omega$
defined by the electron number $N$ and the density parameter $r_\mathrm{s}$.  The latter
defines the radius of a sphere whose volume is equal to
$\Omega/N$~\cite{Martin2004}. Throughout this work, we employ Hartree atomic units.



For the UEG, plane waves are solutions of the Hartree--Fock (HF) Hamiltonian
\begin{equation}
 \phi_{p}(\vec{r}) =\frac{1}{\sqrt{\Omega}} e^{i\vec{k}_{p} \cdot \vec{r} },
\end{equation}
where the wave vectors $\vec{k}_{p}$ represent reciprocal lattice vectors of
the simulation cell. The corresponding HF eigenenergies are
\begin{equation}
\varepsilon_p = \frac{1}{2} \vec{k}_p^2 - \sum_{i \in \mathrm{occ.}} \upsilon_{ip}^{pi},
\label{eq:eigi}
\end{equation}
where $i$ label occupied orbitals. The two\-/electron Coulomb integrals are given by
\begin{equation}
\upsilon_{rs}^{pq} = \upsilon(\vec{q})
  \delta_{\vec{k}_r-\vec{k}_p, \vec{k}_q - \vec{k}_s} \text,
\label{eq:ck}
\end{equation}
with the momentum transfer vector $\vec{q} = \vec{k}_p - \vec{k}_r$ and the
corresponding Coulomb interaction $\upsilon(\vec{q}) = \frac{4\pi}{\Omega
\vec{q}^2}$ for $\vec q\neq \vec 0$. At $\vec q=\vec 0$ the Coulomb interaction
is singular.  However, this singularity is integrable and various techniques
exist to resolve it (see Ref.~\cite{Sundararaman2013} and references therein).
In this work we employ the regularization as described by Fraser et
al.~\cite{Fraser1996}.  An important characteristic of the UEG is the
conservation of momentum.
As evident from Eq.~(\ref{eq:ck}), Coulomb integrals are non\-/zero only if the
momentum transfer vector of the left indices is identical to the negative
transfer vector of the right indices (see Fig.~\ref{fig:rez-mesh}).
%
%
For sufficiently large densities, as
employed in this work, the HF orbital energies are
strictly ordered by length of the corresponding wave vector.

Here and throughout this article we use the following index labels
\begin{center}
\begin{tabular}{ l l }
 $i$, $j$, $k$, $\hdots$  & occupied states,  \\
 $a$, $b$, $c$, $\hdots$  & virtual states in finite basis, \\
 $\alpha$, $\beta$, $\gamma$, $\hdots$ & virtual states beyond finite basis set \\
 & 	referred to as  \emph{augmented virtual states}.\\
\end{tabular}
\end{center}
%
$p$, $q$, $r$ and $s$ are used to label
states that may be both, either occupied or virtual.

We restrict ourselves to the paramagnetic UEG.
Therefore, the number of occupied states $N_\mathrm{o}$ is half the number of electrons
$N$. The momentum of the occupied orbital with the highest eigenenergy is
called the Fermi wave vector, $k_\mathrm{F}$, and the sphere of radius
$k_\mathrm{F}$ is called Fermi sphere.

The number of virtual states $N_\mathrm{v}$ is determined by a plane\-/wave
cutoff momentum $k_\mathrm{cut}$ which is typically much larger than
$k_\mathrm{F}$.  The number of virtual states $N_\mathrm{v}$, contained between
the spheres with radius $k_\mathrm{cut}$ and $k_\mathrm{F}$ is proportional to
$k_\mathrm{cut}^3$~\cite{shepherd2012}.  We stress that in the UEG the orbitals are
unchanged if the number of virtual orbitals is changed (cf.\ the generalized
Brillouin condition in explicit correlated methods~\cite{Haettig2012}).

We refer to the infinite number of plane waves with
momentum exceeding $k_\mathrm{cut}$ in magnitude as
\emph{augmented virtual states}.  These states are considered
theoretically only to better understand and investigate BSIEs of central
quantities and various contributions to the correlation energy.  Consequently,
there appear contributions, for instance Coulomb integrals, involving one
state from the virtual states in the finite basis and another state from the
augmented basis set, viz.\ $\upsilon_{ij}^{a \beta}$.
For a given choice of $a$ and $i$, momentum conservation dictates that the
number of non\-/zero choices for $\beta$ and consequently $j$
is less than the number of occupied states $N_\mathrm{o}$.
This is illustrated in Fig.~\ref{fig:rez-mesh}.
Thus, these contributions will
be negligible in the limit of an infinitely large basis set.

%
%

\subsection{Coupled\-/cluster theory}

In this work we employ single\-/reference coupled\-/cluster theory to approximate
electronic correlation effects. In coupled\-/cluster theory the wave function is given by
\begin{equation}
\label{eq:ccansatz}
       \left| \Psi_\mathrm{cc} \right> = e^{\hat T} \left| \Phi \right>,
\end{equation}
where  $\hat{T}$ is the cluster operator, and $\left|\Phi \right>$ is the reference
wave function. Throughout this work, we use a HF reference determinant. The
full cluster operator for a system with $N$ electrons can be written as
\begin{equation}
\hat{T} = \sum_I^{L} \hat{T}_I \text,
\end{equation}
which is typically truncated at some excitation level $L$.


\subsubsection{Coupled\-/cluster doubles theory}
\label{sec:ccd}

Since all single excitations are
zero~\cite{Shepherd2014} in the UEG, the lowest non\-/zero order coupled\-/cluster ansatz is CCD,
only including double excitations.  The corresponding amplitudes $t_{ij}^{ab}$
which enter the cluster operator $\hat{T}$ in Eq.~(\ref{eq:ccansatz}), are
obtained from the following amplitude equation
\begin{widetext}
\begin{align}
\nonumber
\Delta^{ij}_{ab} & t_{ij}^{ab} =
      \upsilon_{ij}^{ab}
     + \sum_{cd} \upsilon_{cd}^{ab} t_{ij}^{cd}
     + \sum_{kl} t_{kl}^{ab} \sum_{cd} \upsilon_{cd}^{kl} t_{ij}^{cd}
     - P \sum_{k} 2 t_{kj}^{ab} \sum_{cdl} \upsilon_{cd}^{kl} t_{il}^{cd}
     + P \sum_{k} t_{kj}^{ab}   \sum_{cdl} \upsilon_{dc}^{kl} t_{il}^{cd} \\
\nonumber
     &+ \sum_{kl} \upsilon_{ij}^{kl} t_{kl}^{ab}
     - P \sum_{kc} \upsilon_{ci}^{ak} t_{kj}^{cb}
     - P \sum_{kc} \upsilon_{ci}^{bk} t_{kj}^{ac}
     + P \sum_{kc} 2 \upsilon_{ic}^{ak} t_{kj}^{cb}
     - P \sum_{kc} \upsilon_{ic}^{ak} t_{kj}^{bc} \\
\nonumber
     &- P \sum_{c} 2 t_{ij}^{cb} \sum_{dkl} \upsilon_{cd}^{kl} t_{kl}^{ad}
     + P \sum_{c} t_{ij}^{cb}   \sum_{dkl} \upsilon_{dc}^{kl} t_{kl}^{ad}
     + P \sum_{ld} 2 t_{il}^{ad} \sum_{kc} v_{dc}^{lk} t_{kj}^{cb}
     + P \sum_{ld} \frac{1}{2} t_{il}^{da} \sum_{kc} \upsilon_{cd}^{lk} t_{kj}^{cb}   
     + P \sum_{ld} \frac{1}{2} t_{il}^{db} \sum_{kc} \upsilon_{cd}^{lk} t_{kj}^{ac} \\
     &- P \sum_{ld}             t_{il}^{da} \sum_{kc} \upsilon_{dc}^{lk} t_{kj}^{cb}   
     + P \sum_{ld} \frac{1}{2} t_{il}^{da} \sum_{kc} \upsilon_{dc}^{lk} t_{kj}^{bc}  
     - P \sum_{ld}             t_{il}^{ad} \sum_{kc} \upsilon_{dc}^{lk} t_{kj}^{bc}   
     - P \sum_{ld}             t_{il}^{ad} \sum_{kc} \upsilon_{cd}^{lk} t_{kj}^{cb}   
     + P \sum_{ld} \frac{1}{2} t_{il}^{ad} \sum_{kc} \upsilon_{cd}^{lk} t_{kj}^{bc},  
\label{eq:ccd-amplitude}
\end{align}
\end{widetext}

with
\begin{equation}
        P\{\dots\}_{ij}^{ab} =  \{\dots\}_{ij}^{ab} + \{\dots\}_{ji}^{ba} \text.
\end{equation}
The $\Delta^{ij}_{ab}$ (and later $\Delta^{ijk}_{abc}$) terms contain the HF
eigenenergies and are defined by
\begin{equation}
        \Delta^{i \hdots}_{a \hdots} = \varepsilon_i - \varepsilon_a + \hdots\:\text{.}
\end{equation}
The amplitude equation is solved iteratively until a self\-/consistent solution
for the amplitudes $t_{ij}^{ab}$ is found.
Note that it follows from momentum conservation of the Coulomb integrals (see Eq.~(\ref{eq:ck})),
that all amplitudes $t_{ij}^{ab}$ not conserving momentum are zero.
The converged amplitudes can then be
used to evaluate the energy contribution beyond the HF energy, the so\-/called
correlation energy
\begin{equation}
E_\mathrm{c}^\mathrm{D}(t) = \sum_{ijab} \upsilon_{ab}^{ij}
  \left( 2 t_{ij}^{ab} - t_{ji}^{ab} \right) \text.
\label{eq:ccd-energy}
\end{equation}
We want to point out the well\-/known connection between second\-/order Møller--Plesset
perturbation theory (MP2), third\-/order M\o ller--Plesset perturbation theory (MP3),
and CCSD. Using only the first term on the right\-/hand\-/side of Eq.~(\ref{eq:ccd-amplitude}),
one retrieves the MP2 amplitudes, denoted by
\begin{equation}
\label{eq:mp2-amps}
{t^{(1)}}_{ij}^{ab} = \frac{v_{ij}^{ab}}{\Delta^{ij}_{ab}}.
\end{equation}
Evaluating Eq.~(\ref{eq:ccd-energy}) with these amplitudes one obtains the MP2
correlation energy. We stress that for the UEG each element of the MP2
amplitudes is completely described by the HF eigenenergies and the Coulomb
integral. Thus, the elements of ${t^{(1)}}_{ij}^{ab}$ will not change as the basis
set size is increased. The equation for the MP3 amplitudes $t^{(2)}$ can also be inferred
directly from Eq.~(\ref{eq:ccd-amplitude})
by substituting $t^{(2)}$ for $t$ on the left\-/hand\-/side
and $t^{(1)}$ for $t$ for all contributions on the right\-/hand\-/side
that are linear in $t$, while disregarding the terms that are quadratic.
Evaluating Eq.~(\ref{eq:ccd-energy}) with $t^{(2)}$ yields the MP3 correlation
energy. Note that the MP2 and MP3 correlation energies per electron diverge in the thermodynamic limit
$N\rightarrow\infty$~\cite{Shepherd2013}. However, in the present
case, we employ a simulation cell with a finite number of electrons, where finite
order perturbation theories also yield finite correlation energies.

\subsubsection{Triple particle\-/hole excitations}
\label{sec:triples}

The natural extension of CCSD would be the full inclusion of triple particle\-/hole
excitation operators, denoted as CCSDT. This requires the solution
of the corresponding amplitude equations for $t_{ijk}^{abc}$.
However, the storage requirements of these additional terms is $N_\mathrm{o}^3
N_\mathrm{v}^3$, which makes the approach impractical for larger system sizes.
Hence, approximate CCSDT models have been investigated early
on~\cite{Lee1984a,Lee1984b,Urban1985,Noga1987b}. In today's calculations, the
most popular among these methods is the CCSD(T)~\cite{Raghavachari1989}
approach.  Recently, a modified variant, known as the CCSD(cT)
method~\cite{Masios2023}, has been proposed. This approximation includes additional terms
beyond the CCSD(T) method, providing a non\-/diverging description of zero\-/gap
materials in the thermodynamic limit. As single excitations are absent in the UEG system, we
introduce the CCD(T) and CCD(cT) method. The correlation energy beyond $E_\mathrm{c}^\mathrm{D}$
for these methods is given by
\begin{widetext}
\begin{equation}
\label{eq:corr_triples}
        E_\mathrm{c}^\mathrm{(T)} = \sum_{ijk} \sum_{abc}  \left(
          \overline{W}_{ijk}^{abc} + \overline{W}_{ikj}^{acb} +
          \overline{W}_{kji}^{cba} + \overline{W}_{jik}^{bac} +
          \overline{W}_{jki}^{bca} + \overline{W}_{kij}^{cab}
          \right) A_{abc}^{ijk} \text,
\end{equation}
\end{widetext}
with
\begin{equation}
\label{eq:triples_residuum}
  W_{ijk}^{abc} =
                  \sum_e t_{ij}^{ae}\upsilon_{ek}^{bc}
                - \sum_m t_{im}^{ab}\upsilon_{jk}^{mc}
,
\end{equation}
where we define for any six-index quantity $x_{ijk}^{abc}$
\begin{equation}
\label{eq:perm2}
\overline{x}_{ijk}^{abc} = 8 x_{ijk}^{abc}
                         - 4 x_{ijk}^{acb}
                         - 4 x_{ijk}^{cba}
                         - 4 x_{ijk}^{bac}
                         + 2 x_{ijk}^{bca}
                         + 2 x_{ijk}^{cab}
\text.
\end{equation}
The quantity $A$ for the (T) model is given by
\begin{equation}
A_{abc}^{ijk} = \frac{W_{abc}^{ijk}}{\Delta_{abc}^{ijk}}\text.
\end{equation}

In (cT), $A_{abc}^{ijk}$ contains further terms beyond those defined in
Eq.~(\ref{eq:triples_residuum}). The full set of equations for this method is
given in Ref.~\cite{Masios2023}. Here, we provide the terms excluding all
singles contributions. Instead of $W_{ijk}^{abc}$ as defined in
Eq.~(\ref{eq:triples_residuum}), $W_{ijk}^{'abc}$ is used to construct
$A_{abc}^{ijk}$, which is defined as
\begin{equation}
\label{eq:ct_residuum}
W_{abc}^{'ijk} =  
      \sum_e t_{ij}^{ae} J_{ek}^{bc} - \sum_m t_{im}^{ab} J_{jk}^{mc}
,
\end{equation}
with
\begin{widetext}
\begin{align}
J^{bc}_{ek} &= \upsilon^{bc}_{ek} + \sum_{mn} t_{mn}^{cb}\upsilon_{ke}^{mn}
            + \sum_{mf} \left( 2\upsilon_{ef}^{bm} t_{km}^{cf} -  \upsilon_{ef}^{bm} t_{km}^{fc}
                                - \upsilon_{fe}^{bm}t_{mk}^{fc} -  t_{km}^{fb}\upsilon_{fe}^{cm} \right) \text,
\label{eq:Jp}
             \\
J^{mc}_{jk} &= \upsilon_{jk}^{mc}  + \sum_{ef} t_{kj}^{ef}\upsilon_{ef}^{cm}
               + \sum_{nf} \left( 2\upsilon_{jf}^{mn}t_{kn}^{cf} - \upsilon_{jf}^{mn}t_{kn}^{fc}-
              \upsilon_{jf}^{nm}t_{nk}^{fc} - t_{nj}^{cf}\upsilon_{kf}^{nm} \right) \text.
\label{eq:Jh}
\end{align}
\end{widetext}

We emphasize that the large benefit of both approaches is the inclusion of
triply excited clusters without storing an intermediate quantity of the size
$N_\mathrm{v}^3 N_\mathrm{o}^3$. Nevertheless, the memory footprint of (cT) is
roughly doubled compared to the (T) approach.  This is because $J_{ek}^{bc}$ is
typically computed once and stored in memory. The computational cost for the
evaluation of $J_{ek}^{bc}$ and $J_{jk}^{mc}$ is negligible compared to the
contractions in Eqs.~(\ref{eq:triples_residuum}) and~(\ref{eq:ct_residuum}). Still,
the total number of operations in (cT) is approximately twice that of the operations in a
(T) calculation. In (T), we need to evaluate Eq.~(\ref{eq:triples_residuum}),
which is the rate determining contraction scaling as $\mathcal{O}(N^7)$. For
(cT), however, one has to evaluate both the contractions in
Eq.~(\ref{eq:triples_residuum}) and Eq.~(\ref{eq:ct_residuum}).

\section{Computational details}
\label{sec:compDetails}

All UEG MP2, CCD, CCD(T), and CCD(cT) calculations have been performed using a
recently developed code~\cite{cc4ueg}. This code fully employs the momentum
conservation of the Coulomb integrals and amplitudes, resulting in a reduction
of the storage requirements for CCD calculations from $N^4$ to $N^3$.
Additionally, the number of operations in the CCD equations decreases from
$N^6$ to $N^4$. Fully converged CCD amplitudes are obtained by solving the
amplitudes equation Eq.~(\ref{eq:ccd-amplitude}) iteratively. We found an
energy criteria of $10^{-8}$~a.u. sufficient for the analysis performed here.

We stress that the presented results have only a weak dependence on the
number of electrons in the unit cell. A larger electron number reduces the
so\-/called finite\-/size error with respect to the thermodynamic limit. This
error does not strongly interfere with the investigated BSIE. Importantly,
the power\-/laws of the BSIE discussed here are fundamental and independent of the
electron number. For the CCD analysis, we will work with a density of
$r_\mathrm{s}=5.0$~a.u. with 54 electrons, while for the analysis of the triple excitations,
with 14 electrons. The influence of the electron number and density will be further analyzed
in Sections~\ref{sec:dep_num_den} and \ref{sec:dep_num_tri}.

\section{Large\-/momentum limit results for various CCD theories}
\label{sec:ccdbsie}

We now turn to the discussion of the obtained results at the level of
(approximate) CCD theory.


The BSIE originates from truncating the number of virtual states. In the UEG
system, virtual states with high kinetic energy carry a momentum which is large
compared to the Fermi sphere radius defining the set of occupied orbitals. This
implies that for the two virtual indices in Eq.~(\ref{eq:mp2-amps}),
$\vec{q} \approx \vec{k}_\alpha \approx -\vec{k}_\beta$. In this limit, the denominator
of Eq.~(\ref{eq:mp2-amps}) is dominated by the kinetic energy contribution of
the virtual states, and $\Delta_{\alpha \beta}^{ij} \propto q^2$, where $q$ always
represents $|\vec q|$. The Coulomb integral $\upsilon_{ij}^{\alpha \beta}$ becomes
proportional to $q^{-2}$, leading to an asymptotic behavior of
${t^{(1)}}_{ij}^{\alpha \beta} \propto q^{-4}$. It is straightforward to assign a
transfer momentum $\vec{q}$ to a Coulomb integral $\upsilon_{ij}^{\alpha \beta}$ [see
Eq.~(\ref{eq:ck})] and therewith to an amplitude $t_{ij}^{\alpha \beta}$.


When we increase the number of virtual states, i.e. enlarge the radius
$k_\mathrm{cut}$,  the number of accessible transfer vectors $\vec{q}$ increase
accordingly. This implies for the energy expression in Eq.~(\ref{eq:ccd-energy}),
$\lim_{q \to \infty} \int_q^{\infty} dq' q'^2 \frac{1}{q'^2} \frac{1}{q'^4}
\propto q^{-3}$, which is in accordance with the well known $N_\mathrm{v}^{-1}$
convergence behavior of the correlation energy.

The BSIE $\Delta E$ is defined as the difference between the energy obtained
from a calculation with a finite virtual basis set and the estimate from the
complete basis set (CBS). In this work, CBS estimates are obtained by extrapolating
energies from the two largest basis sets used for the given
system, employing the corresponding power law.

\subsection{Diagrammatic contributions to the CCD energy}

\begin{figure}[t]
\includegraphics{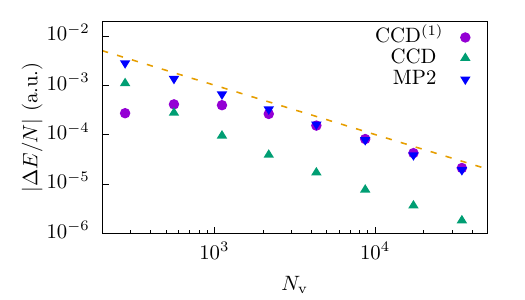}
\caption {
The plot displays the BSIE $|\Delta E|$ per electron for
MP2, CCD$^{(1)}$, and CCD.  The dashed line is proportional to $N_\mathrm{v}^{-1}$.
Results are shown for the 54 electron system at a density $r_\mathrm{s}=5.0$~a.u..
CBS estimates are obtained by $N_\mathrm{v}^{-1}$ extrapolation using the two
largest systems.
}
 \label{fig:mp_ccd_energy}
\end{figure}

In the present work, we partition the correlation energy and related
quantities according to the right\-/hand\-/side contributions in
Eq.~(\ref{eq:ccd-amplitude}). We refer to each contribution as a \emph{channel},
aiming to identify distinct large\-/momentum behaviors in different channels, in
order to improve or even justify correction schemes for the BSIE.
To this end, we introduce \emph{channel amplitudes}, denoted as $t^{(X)}$ and defined by
the expression
\begin{equation}
  {t^{(X)}}^{ab}_{ij}(t) = \frac{%
    \textnormal{Term $(X)$ of rhs.~of Eq.~\ref{eq:ccd-amplitude} with given $t$}
  }{\Delta^{ij}_{ab}},
  \label{eq:channel_amplitudes}
\end{equation}
where $X$ represents one of the terms on the right\-/hand\-/side of
Eq.~(\ref{eq:ccd-amplitude}). For example, the MP2 and
the PPL channel amplitudes stem from the
first and the second term on the right\-/hand\-/side of Eq.~(\ref{eq:ccd-amplitude}),
denoted by (a) and (b), respectively. They can be expressed as
${t^{(\textrm{a})}}^{ab}_{ij} = \upsilon^{ab}_{ij}/\Delta^{ij}_{ab}$ and
${t^{(\textrm{b})}}^{ab}_{ij} = (
  \sum_{cd}\upsilon^{ab}_{cd} t^{cd}_{ij}
)/\Delta^{ij}_{ab}$.
The channel amplitudes depend on the choice of approximation for the
doubles amplitudes $t$ on the right\-/hand\-/side of Eq.~(\ref{eq:ccd-amplitude}).
We examine three cases: (i) $t=0$, (ii) $t=t^{(1)}$, and (iii) the $t$ that is
the fully self\-/consistent solution of Eq.~(\ref{eq:ccd-amplitude}).
In case (i), only the MP2 channel (a) is non\-/zero yielding
the MP2 amplitudes $t^{(\textrm{a})}(0)=t^{(1)}$.
In the cases (ii) and (iii), all channels $t^{(X)}(t)$ are non\-/zero and
depend on the argument amplitudes $t$. The first channel (a) always gives
the MP2 amplitudes, as it is independent on $t$.
We label the results obtained for the cases (i), (ii), and (iii) as
MP2, CCD$^{(1)}$, and CCD, respectively.
While a similar analysis was performed in a previous work for some
contributions~\cite{Irmler2019}, our current study extends beyond the prior work.
We also note that the contributions to CCD$^{(1)}$ that are linear in $t^{(1)}$
are identical to MP3.

Results for the density corresponding to
$r_\mathrm{s}=5.0$~a.u. are presented in Fig.~\ref{fig:mp_ccd_energy}. In all
calculations, the BSIE converges like $N_\mathrm{v}^{-1}$.
However, we notice that the magnitude of $\Delta E$ for a given basis set
differs significantly between MP2, CCD$^{(1)}$ and CCD.
Moreover, we note that $\Delta E$ approaches zero in the CBS limit with an
opposite sign in CCD$^{(1)}$ compared to MP2 and CCD.
Consequently, the following question arises: \emph{which contributions are
responsible for these differences?}
%

\begin{widetext}

\begin{figure}[H]
\includegraphics{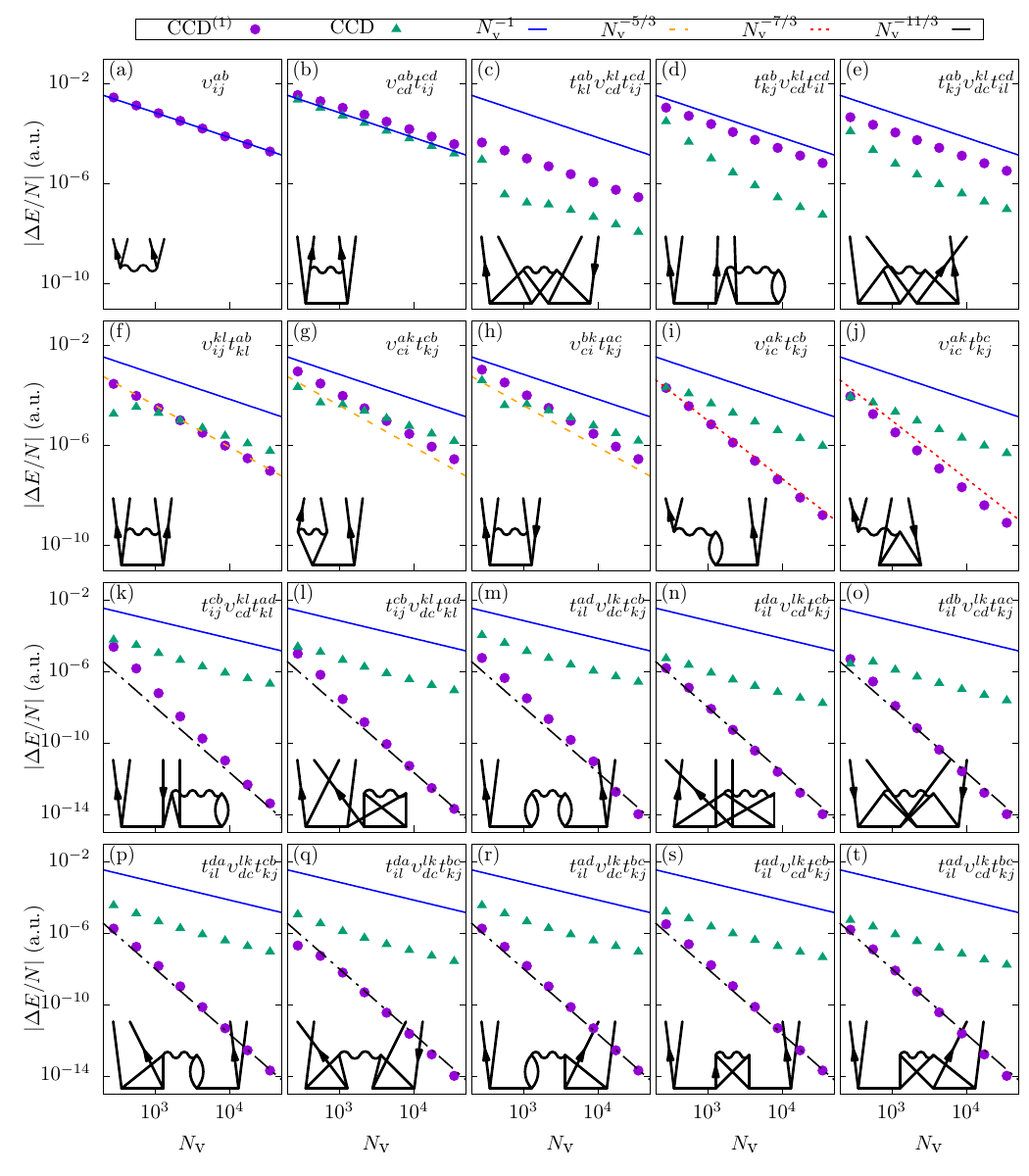}
\caption {Shown are the absolute BSIEs $|\Delta E|$ per electron of all
contributions to the CCD and CCD$^{(1)}$ correlation energy expressions. The
individual terms are given in order of appearance in
Eq.~(\ref{eq:ccd-amplitude}). In order to simplify the assignment we provide
the corresponding equations (prefactors and sums are suppressed), as well as
their diagrammatic representation. The lines of various color, indicating the
different power laws, are identical in all plots.  Results are shown for a
system with 54 electrons and $r_\mathrm{s} = 5.0$~a.u..  CBS estimates are
obtained from extrapolation of the two largest systems using the corresponding
power law.}
   \label{fig:energy_channels}
\end{figure}
\end{widetext}

We now turn to a detailed analysis of the BSIE and the rate of convergence for
all diagrammatic channels ${t^{(X)}}(t)$. These channels are computed from the
given amplitudes $t$, which are either the MP2 amplitudes $t^{(1)}$ or the solution of the
amplitude equation [Eq.~(\ref{eq:ccd-amplitude})], referred to as CCD$^{(1)}$ and CCD,
respectively.  Due to the large number of terms, this is a complex and
elaborate endeavor.  However, using a numerical approach one can readily
obtain all BSIEs using CCD$^{(1)}$ and CCD theory.
In Fig.~\ref{fig:energy_channels}, we illustrate all individual BSIEs through 20 plots
labeled (a)--(t). The organization of these plots is as follows:
Figs.~\ref{fig:energy_channels}(a)--(e) depict the BSIEs of all terms
exhibiting a convergence of $N_\mathrm{v}^{-1}$ at the level of CCD$^{(1)}$
theory. Since MP2 is contained in both CCD$^{(1)}$ and CCD,
Fig.~\ref{fig:energy_channels}(a) is identical to MP2 from
Fig.~\ref{fig:mp_ccd_energy}. For Figs.~\ref{fig:energy_channels}(f)--(h),
(i)--(j), and (k)--(t) the CCD$^{(1)}$ BSIEs exhibit convergence rates of
$N_\mathrm{v}^{-5/3}$, $N_\mathrm{v}^{-7/3}$, and $N_\mathrm{v}^{-11/3}$,
respectively.  In contrast, all CCD BSIEs converge as $N_\mathrm{v}^{-1}$.
The results depicted in Fig.~\ref{fig:energy_channels} show that the
PPL term in Fig.~\ref{fig:energy_channels}(b)) is by far the most important
contribution, besides the MP2 term, shown in Fig.~\ref{fig:energy_channels}(a).
Consequently, our analysis begins with the PPL contribution. Because we
are only interested in the fundamental power laws, signs and prefactors will be
suppressed in the following analysis.

\subsubsection{The particle--particle ladder contribution}
\label{sec:ppl}

We now discuss the basis set convergence of the  PPL contribution to the
CCD$^{(1)}$ and CCD energy. As depicted in Fig.~\ref{fig:energy_channels}(b), this
contribution converges in both approaches at the same rate as the MP2 term,
scaling as $N_\mathrm{v}^{-1}$. The magnitude of the PPL contribution is found to be comparable
to that of the MP2, and especially in the case of CCD$^{(1)}$, the PPL contribution
is even larger than the MP2 term. These results have been obtained for the same density and number
of electrons as in Fig.~\ref{fig:mp_ccd_energy}.  We have already discussed the
significance of this contribution to the BSIE of CCSD theory in
Refs.~\cite{Irmler2019,Irmler2019a,Irmler2021}, where different approaches have
been presented to account for the BSIE of the PPL contribution. In
Refs.~\cite{Irmler2019,Irmler2019a}, we also provide explanations for its
$N_\mathrm{v}^{-1}$ convergence rate. Here, we briefly reiterate the
explanations by splitting the contributions to the amplitudes into the
conventional and augmented virtual basis set. We suppress all contributions
which contain amplitudes with one orbital in the finite- and the other in the
augmented virtual basis set as these contributions are negligible [see
Fig.~\ref{fig:rez-mesh}].
\begin{align}
\label{eq:b_conventional}
{t^{(\textrm{b})}}^{ab}_{ij} = \frac{1}{\Delta^{ij}_{ab}}
  \left(
  \sum_{cd}\upsilon^{ab}_{cd} t^{cd}_{ij}
+ \sum_{\gamma \delta}\upsilon^{ab}_{\gamma \delta} t^{\gamma \delta}_{ij}
  \right),\\
\label{eq:b_augmented}
{t^{(\textrm{b})}}^{\alpha \beta}_{ij} = \frac{1}{\Delta^{ij}_{\alpha \beta}}
   \left(
   \sum_{cd}\upsilon^{\alpha \beta}_{cd} t^{cd}_{ij}
 + \sum_{\gamma \delta}\upsilon^{\alpha \beta}_{\gamma \delta} t^{\gamma \delta}_{ij}
   \right).
\end{align}
The first term in the parenthesis of Eq.~(\ref{eq:b_conventional}) is the
PPL contribution using the conventional finite basis. The second term in
Eq.~(\ref{eq:b_conventional}) reveals how amplitude elements from the augmented
virtual basis couple to the amplitudes in the finite virtual basis set.
%
In the UEG, we can approximate the appearing Coulomb interaction
$\upsilon^{ab}_{\gamma \delta}$ in the following way
\begin{equation}
\upsilon(\vec{k}_\gamma - \vec{k}_a)
\approx
\upsilon(\vec{k}_\gamma - \vec{k}_i)
\text.
\label{eq:approxv}
\end{equation}
This is well justified for high lying augmented virtual states with very large
momentum $|\vec{k}_\gamma| \gg |\vec{k}_a| > | \vec{k}_i|$. Using this
approximation, the last term in Eq.~(\ref{eq:b_conventional}) can be written as
$1/\Delta_{ab}^{ij} \sum_{\gamma \delta} \upsilon_{\gamma \delta}^{ij}
t_{ij}^{\gamma \delta}$.  As discussed earlier, this term exhibits a
$1/N_\mathrm{v}$ convergence, indicating that the energy contribution derived from
these amplitudes follows the same $1/N_\mathrm{v}$ convergence behavior.
%
%
Moving on to the first term in Eq.~(\ref{eq:b_augmented}), we can employ the
approximation introduced in Eq.~(\ref{eq:approxv}). This allows us to write the
term as $\sum_{\alpha \beta} \upsilon_{ij}^{\alpha \beta}/\Delta^{ij}_{\alpha
\beta} (\sum_{cd} t_{ij}^{cd}) $. Carrying out the sum in the parenthesis will
lead to a scalar number for each electron pair $ij$. 
This analysis reveals that for momenta $q = | \vec{k}_\alpha - \vec{k}_i | \gg
k_\mathrm{cut}$ the amplitudes stemming from Eq.~(\ref{eq:b_augmented}) become
proportional to those from the MP2 level of theory, importantly, with an
opposite sign.

It is straightforward to see that the other term in Eq.~(\ref{eq:b_augmented})
converges much faster, specifically as $q^{-6}$, and its contribution is
therefore not of leading order in $q$.
We stress that due to the symmetry of the particle-particle term, the energy
contribution from the second term in Eq.~(\ref{eq:b_conventional}) is identical
to the first term in Eq.~(\ref{eq:b_augmented}) at MP3 level of theory. In CCD,
however, this is not necessarily the case.

\subsubsection{Slowly converging quadratic contributions}

We now seek to analyze the three contributions shown in
Fig.~\ref{fig:energy_channels}(c)--(e).  They are all quadratic in amplitude
$t$.  Interestingly, these are the only contributions, beyond MP2 and PPL,
showing a $N_\mathrm{v}^{-1}$ convergence of the BSIE at CCD$^{(1)}$ level
of theory.

To explain this observation, we consider a two electron singlet system and split
the contribution to the amplitudes into the conventional and augmented virtual
basis sets.  The three considered terms here are the third, forth, and fifth
term on the right\-/hand\-/side of Eq.~(\ref{eq:ccd-amplitude}). For the two
electron system, these terms become identical, apart from different prefactors
and read
\begin{align}
{t^{(\textrm{c--e})}}_{ii}^{ab}(t) &= \frac{1}{\Delta^{ii}_{ab}} t_{ii}^{ab}
\left(
              \sum_{cd}  \upsilon_{cd}^{ii} t_{ii}^{cd}
 +            \sum_{\gamma \delta}
              \upsilon_{\gamma \delta}^{ii} t_{ii}^{\gamma \delta}
\right),
\label{eq:cde_conventional}
\\
{t^{(\textrm{c--e})}}_{ii}^{\alpha \beta}(t) &= \frac{1}{\Delta^{ii}_{\alpha \beta}}
             t_{ii}^{\alpha \beta}
\left(
            \sum_{cd}  \upsilon_{cd}^{ii} t_{ii}^{cd}
 +          \sum_{\gamma \delta}
              \upsilon_{\gamma \delta}^{ii} t_{ii}^{\gamma \delta}
\right).
\label{eq:cde_augmented}
\end{align}
The right term in the parenthesis of Eq.~(\ref{eq:cde_conventional}) shows how
the presence of the augmented basis alters the amplitudes belonging to the
conventional basis set. The respective amplitudes are scaled by the terms in
parenthesis. For the two electron singlet, this sum corresponds to the BSIE of
the correlation energy of the electron pair, which converges slowly as
$N_\mathrm{v}^{-1}$. Thus, the quadratic contributions (c--e) exhibit the same
power\-/law in both, CCD$^{(1)}$ and CCD calculations. The amplitudes from the
augmented basis set of Eq.~(\ref{eq:cde_augmented}) show a faster convergence.
As both orbitals $\phi_\alpha$ and $\phi_\beta$ are virtual states with large
momenta, it follows that $\Delta^{ii}_{\alpha\beta} \propto q^2$ and
$t_{ii}^{\alpha\beta} \propto q^{-4}$. This results in an overall
$q^{-6}$ convergence of the corresponding amplitudes. Applying these
amplitudes with augmented virtual states in the energy expression
Eq.~(\ref{eq:ccd-energy}) leads to a $N_\mathrm{v}^{-5/3}$ behavior of the BSIE.

Now we are in the position to explain the significant difference in magnitude
of the BSIE for these contributions in CCD$^{(1)}$ and CCD [see
Fig.~\ref{fig:energy_channels}(c)-(e)].  In the previous section, we have seen
that the amplitude elements of the augmented virtual manifold are altered by
the PPL contribution [see Eq.~(\ref{eq:b_augmented})]. For the studied system
with 54 electrons at a density of $r_\mathrm{s}=5.0$~a.u., the amplitude
elements from the augmented virtual states are significantly smaller in CCD,
than they are in CCD$^{(1)}$.  The predominant BSIE contribution stems from
Eq.~(\ref{eq:cde_conventional}) which contains a contraction of such amplitude
elemts from augmented virtual states.  This explains why the BSIE of the here
discussed terms is larger in CCD$^{(1)}$ compared to CCD.
%
%
%
%

We stress that this analysis, restricting to a two electron singlet, is
evidently limited. The results for the 54 electron system in
Fig.~\ref{fig:energy_channels}(c)--(e) reveal differences of more than one
order of magnitude between the different terms.  However, for the two electron
singlet, all terms are identical, apart from a factor of two.
%

\subsubsection{Other ladder diagrams}
\label{sec:otherLadder}

This section addresses the three other ladder terms,
linear in the amplitude $t$, shown in
Fig.~\ref{fig:energy_channels}(f)--(h). For all three terms, the BSIE converges
as $N_\mathrm{v}^{-5/3}$ in the CCD$^{(1)}$ calculations. We pick one of the
terms, specifically a particle--hole term ${t^{(\mathrm{g})}}^{ab}_{ij}(t)$
depicted in Fig.~\ref{fig:energy_channels}(g),
%
%
to study the origin of this convergence behavior. The remaining two terms can
be treated analogously. We list all contributions after partitioning the
virtual states
into conventional and augmented virtual basis sets:
%
\begin{align}
\label{eq:phzz}
{t^{(\mathrm{g})}}_{ij}^{ab}          &= \frac{1}{\Delta^{ij}_{ab}} \sum_{ck}
               \upsilon_{ci}^{a k} t_{kj}^{cb} + \frac{1}{\Delta^{ij}_{ab}} \sum_{\gamma k}
               \upsilon_{\gamma i}^{a k} t_{kj}^{\gamma b}, \\
\label{eq:phz}
{t^{(\mathrm{g})}}_{ij}^{\alpha b} &= \frac{1}{\Delta^{ij}_{\alpha  b}} \sum_{ck}
               \upsilon_{ci}^{\alpha k} t_{kj}^{c b}
                      + \frac{1}{\Delta^{ij}_{\alpha b}} \sum_{\gamma k}
               \upsilon_{\gamma i}^{\alpha k} t_{kj}^{\gamma  b}, \\
\label{eq:phy}
{t^{(\mathrm{g})}}_{ij}^{a \beta} &= \frac{1}{\Delta^{ij}_{a \beta}} \sum_{ck}
               \upsilon_{ci}^{a k} t_{kj}^{c\beta}
                      + \frac{1}{\Delta^{ij}_{a \beta}} \sum_{\gamma k}
               \upsilon_{\gamma i}^{a k} t_{kj}^{\gamma \beta}, \\
\label{eq:phx}
{t^{(\mathrm{g})}}_{ij}^{\alpha \beta} &= \frac{1}{\Delta^{ij}_{\alpha\beta}} \sum_{ck}
               \upsilon_{ci}^{\alpha k} t_{kj}^{c\beta}
                      + \frac{1}{\Delta^{ij}_{\alpha\beta}} \sum_{\gamma k}
               \upsilon_{\gamma i}^{\alpha k} t_{kj}^{\gamma \beta}.
\end{align}
The first term on the right hand side of Eq.~(\ref{eq:phzz}) is the conventional
expression in the finite basis set. The second term in Eq.~(\ref{eq:phzz}), as
well as the terms in Eqs.~(\ref{eq:phz}) and~(\ref{eq:phy}) contain amplitudes or
Coulomb integrals with one virtual orbital in the finite basis and the other
in the augmented basis set.
The vast majority of these terms are zero due to momentum conservation [see
Fig.~\ref{fig:rez-mesh}]. Non\-/zero contributions can only be found in a
volume corresponding to the Fermi sphere. As a consequence, these
contributions are expected to be negligible compared to the terms in
Eq.~(\ref{eq:phx}). Here, the first term has again one virtual state
from the finite basis and the other from the augmented virtual states. Again,
this contribution is negligible. The second term in Eq.~(\ref{eq:phx})
converges as $q^{-6}$.  This originates from the energy denominator
$1/\Delta^{ij}_{\alpha \beta}$ and the amplitudes $t_{kj}^{\gamma \beta}$
scaling as $q^{-2}$ and $q^{-4}$, respectively. The maximum momentum transfer
in the appearing Coulomb interaction cannot exceed $2k_\mathrm{F}$. Thus, the
Coulomb interaction does not induce a further power of $q^{-2}$.
Additionally, the sum over the states $k$ and $\gamma$ does not alter the
asymptotic behavior, as the number of states fulfilling the momentum
conservation in the Coulomb integral is proportional to $k_\mathrm{F}$. In conclusion,
the three ladder terms discussed here converge as $q^{-6}$, corresponding to $N_\mathrm{v}^{-5/3}$.

Interestingly, the BSIE behavior of ladder terms (f)--(h) changes fundamentally
from $N_\mathrm{v}^{-5/3}$ in CCD$^{(1)}$ to  $N_\mathrm{v}^{-1}$ in CCD
calculations.  The origin of this effect lies in the PPL contribution, and to a
smaller extent, in the other three quadratic contributions discussed in the
previous section. Expressions such as Eqs.~(\ref{eq:b_conventional}) and~(\ref{eq:cde_conventional})
couple the amplitudes of the augmented virtual states
to the amplitudes of conventional virtual states. Consequently, the argument
amplitudes in Eq.~(\ref{eq:channel_amplitudes}) are altered, leading to a change in the energy
contribution of the given channel.

\subsubsection{Linear ring type diagrams}
\label{sec:linearRing}

The terms depicted in Fig.~\ref{fig:energy_channels}(i)--(j) are commonly
referred to as ring and crossed ring diagrams. Similar to the previously discussed
ladder diagrams, the ring and crossed ring diagrams also exhibit a $N_\mathrm{v}^{-1}$
behavior in CCD, arising from the same underlying mechanism.
In CCD$^{(1)}$, however, the BSIE converges as $N_\mathrm{v}^{-7/3}$.
We present the ring contribution as an example, Fig.~\ref{fig:energy_channels}(i),
\begin{equation}
{t^\mathrm{(i)}}_{ij}^{\alpha \beta} = \frac{1}{\Delta^{ij}_{\alpha \beta}}
   \sum_{k\gamma} \upsilon_{i\gamma}^{\alpha k} t_{kj}^{\gamma \beta}
\text,
\end{equation}
which, in the limit of high\-/lying virtual states, exhibits $q^{-2}$,
$q^{-2}$, and $q^{-4}$ contributions, from $\frac{1}{\Delta^{ij}_{ab}}$,
$\upsilon_{ic}^{ak}$, and $t_{kj}^{cb}$, respectively.  This results in
an overall $q^{-8}$ convergence of the amplitudes. In passing, we note that both
diagrams in Fig.~\ref{fig:energy_channels}(i)--(j) become identical in the
large $q$ regime, except for a different prefactor, while it is well\-/known
that they behave very different in the small $q$ limit~\cite{mattuck}.

\subsubsection{Other quadratic contributions}
\label{sec:otherQuadratic}

The remaining ten contributions in Fig.~\ref{fig:energy_channels}(k)--(t) are
all quadratic in the amplitudes $t$. The BSIE of these contributions shows a fast
decay in CCD$^{(1)}$, scaling as $N_\mathrm{v}^{-11/3}$.
As before, this behavior can be explained by
the aggregation of factors $q^{-4}$ for each amplitude, the factor $q^{-2}$
once for the Coulomb interaction mediating the momentum $q$ and once
for the energy denominator $1/\Delta_{\alpha \beta}^{ij}$. Also here,
the convergence behavior of the BSIE changes in the CCD calculation and
becomes inversely proportional to the number of employed states, $N_\mathrm{v}$.

\subsection{Structure Factor analysis}

The static structure factor is a pivotal quantity in the context of periodic
electronic structure theory~\cite{Martin2016}.  In coupled\-/cluster
approaches, a related quantity, the so\-/called transition structure factor
$S$, was employed in a number of recent
works~\cite{Liao2016,Gruber2018,Mihm2023}. The structure factor is directly
related to the correlation energy contribution at a given  momentum transfer
$\vec{q}$, according to the following equation
\begin{equation}
E_\mathrm{c} = \sum_{\vec q} \upsilon(\vec{q}) S(\vec{q}).
\label{eq:sf}
\end{equation}
For energy expressions in the form of Eq.~(\ref{eq:ccd-energy}), the transition
structure factor can be written as
\begin{equation}
S(\vec{q}) = \sum_{ijab} \left(2 t_{ij}^{ab} - t_{ji}^{ab} \right)
  \delta_{\vec{q},\vec{k}_a-\vec{k}_i} \delta_{\vec{q},\vec{k}_j-\vec{k}_b},
\label{eq:definesf}
\end{equation}
where the amplitudes $t^{ab}_{ij}$ have been obtained from calculations with
a given finite virtual basis set.
\begin{figure}
\includegraphics{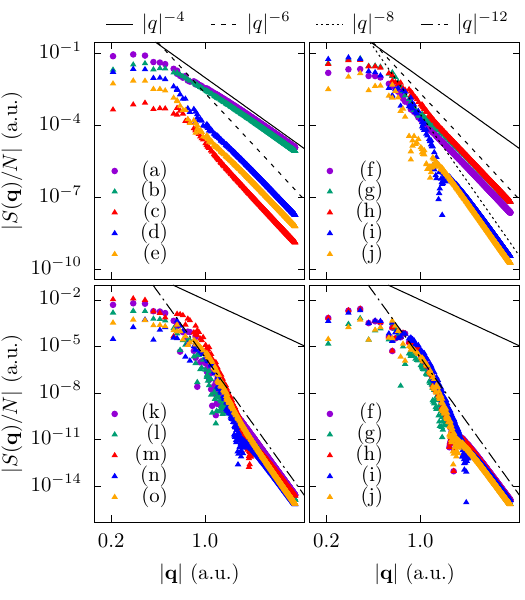}
\caption { Transition structure factor results for the individual diagrammatic
channels. Results are obtained from a converged CCD calculation with 54
electrons at $r_\mathrm{s} = 5.0$~a.u. and 67664 virtual orbitals. All figures show
various diagrammatic contributions, labeled as in
Fig.~\ref{fig:energy_channels}. In addition, four lines with different powers
of $q^{-n}$ are shown. The $q^{-4}$ curve is the same in all plots and allows
comparison between the four different panels.}
 \label{fig:sf_ccd}
\end{figure}

As we have elaborated in previous sections, in CCD the elements of the
amplitudes depend on the employed basis set. Likewise, this holds for
the transition structure factor, as indicated by Eq.~(\ref{eq:definesf}).
Consequently, we analyze the transition structure factor using a large basis
set with more than 2500 virtual states per occupied orbital. We do not analyze
results from CCD$^{(1)}$ calculations but restrict to results from fully
converged CCD amplitudes. The channel\-/resolved transition structure factor
$S^{(X)}$ can be defined by employing the respective channel amplitudes
$t^{(X)}$ in Eq.~(\ref{eq:definesf}).
Fig.~\ref{fig:sf_ccd} displays the transition structure factors in the limit of
large transfer momenta $\vec{q}$. This allows us to draw conclusions about the
features at very short interelectronic distances. Interestingly, we identify
only two channels that are of leading order for large $\vec{q}$, which is a
central finding of the present work. These contributions correspond to the MP2
and the PPL contribution, both showing a $q^{-4}$ decay for large values of
$\vec{q}$.  It follows from Eq.~(\ref{eq:sf}) that this $q^{-4}$ behavior of the
transition structure factor corresponds to a $N_\mathrm{v}^{-1}$ convergence of
the BSIE.  This implies that all other channels do not significantly alter the
transition structure factor at large momentum transfers $\vec{q}$.
Consequently, they cannot change the linear slope of the singlet transition
pair correlation function at the coalescence point. We emphasize that the
transition pair correlation function is not an observable. However, these
results corroborate that, for all channels besides MP2 and PPL, the slow
$N_\mathrm{v}^{-1}$ convergence of the BSIE actually arises from long\-/range
modulations of the corresponding channel decomposed transition structure.

\subsection{Dependence on electron number and density}


\label{sec:dep_num_den}
\begin{table}
\caption{Shown is the ratio of BSIEs between different channels and the MP2
term. These ratios are obtained from calculations with sufficiently large basis
sets ($N_\mathrm{v}/N_\mathrm{o} \approx 150-200$). We have grouped the
slowly quadratic contributions, i.e. (c)--(e), the linear ladder terms except
PPL, (f)--(j), and finally all other quadratic terms (k)--(t). All ratios are
scaled by $10^{-3}$.  }
\frame{
\begin{tabular}{  c | r r r | r r r }
& \multicolumn{3}{c}{$r_\mathrm{s}=1.0$} & \multicolumn{3}{c}{$r_\mathrm{s}=5.0$} \\
$\Delta E^{(X)}/\Delta E^\mathrm{(\mathrm{a})}$& 14 & 54 & 162 & 14 & 54 & 162 \\
\hline

(b)     & $-410$ & $-400$ & $-404$ & $-798$ & $-795$ & $-797$ \\
(c)--(e) & $-11$ & $-12$ & $-8.3$ & $ 0.9$ & $-1.6$ & $-3.7$ \\
(f)--(j) & $-24$ & $-15$ & $-24$ & $-97$ & $-87$ & $-80$ \\
(k)--(t) & $-0.5$ & $-2.5$ & $-1.5$ & $-1.2$ & $-0.3$ & $-0.9$ \\
\end{tabular}
}
\label{tab:bsieEl}
\end{table}

In the previous sections, we carefully studied the BSIE of different
contributions to the CCD energy contribution. We conducted the investigations
for a single system of 54 electrons at the density $r_\mathrm{s}=5.0$~a.u.. We saw
that other than the MP2 term, the PPL contribution has by far the largest BSIE.
As the BSIE of PPL and MP2 have opposite signs, this can lead to a
significant decrease of the overall BSIE in CCD.

Here, we show how these main findings hold for different electron numbers and
densities, respectively. However, we are not analyzing CCD$^{(1)}$ but restrict
the analysis to the CCD level of theory, which is the main interest of the
present work. We analyze the BSIE of the different diagrammatic channels,
using the fact that in CCD all channels show the same $N_\mathrm{v}$ decay.
Table~\ref{tab:bsieEl} shows results for three different electron numbers and
two different electron densities, respectively.
For the high density system, $r_\mathrm{s}=1$, all terms except the PPL terms
show a very small contribution to the total BSIE. Only the PPL term has a
significant BSIE compared to MP2, which is around 40\%.  The PPL term is
virtually independent of the system size for both densities considered.

In the low density system, at $r_\mathrm{s}=5.0$~a.u., all quadratic contributions,
i.e.\ (c)--(e) and (k)--(t), show a small BSIE.  The BSIE of all linear terms
(f)--(j) is significantly larger compared to the high density system. The BSIE
of the PPL contribution is around 80\% of the MP2 value. Also the other linear
terms in (f)--(j) are much more pronounced at low densities, with a BSIE approximately
10\% of the BSIE from MP2. As for the other density, these results
show no strong dependence on the employed electron number.
We emphasize that the PPL contribution becomes more important in comparison
to MP2 theory in the low density limit. This is expected because higher order
perturbation theory terms become more significant as $r_\mathrm{s}$ increases.
Furthermore, this implies that any truncated finite\-/order perturbation theory
approach, such as CCD$^{(1)}$, becomes unreliable. Indeed we find that
CCD$^{(1)}$ overestimates the PPL contribution compared to CCD significantly.
However, similar to the resummation over ring diagrams, CCD performs a
resummation over ladder diagrams, which is extremely important for a well
balanced estimate of the PPL contribution in the low density limit.

\section{Triple excitations}
\label{sec:tbsie}

This section is dedicated to the BSIE of the perturbative triples contributions
presented in section~\ref{sec:triples}. We start the analysis with the (T)
method and later discuss the differences for the (cT) method.

\subsection{The BSIE of the (T) model}

\begin{figure}
  \includegraphics{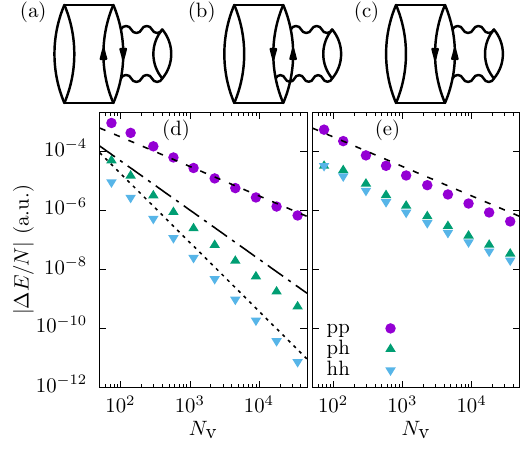}
  \label{fig:t_sub2}
\caption { Figures (a), (b), and (c) diagrammatically illustrate the three
different contributions hh, ph, and pp, respectively. (d) and (e) display the
BSIE per electron of the three mentioned contributions for a system with 14
electrons and $r_\mathrm{s}=5$~a.u..  In (d), MP2 amplitudes are employed,
while (e) shows results with CCD amplitudes.  CBS estimates are obtained from
extrapolation of the two largest systems using the corresponding power law.
Three different lines in (b) show the different power laws, namely $N^{-1}$,
$N^{-5/3}$, and $N^{-7/3}$.}
\label{fig:t3}
\end{figure}

The energy expression of the (T) contribution in Eq.~(\ref{eq:corr_triples})
contains two different contractions, namely a particle contraction and a hole
contraction [Eq.~(\ref{eq:triples_residuum})]. As a consequence, the total
energy expression can be split into three terms. One term which contains only
hole contractions, denoted as (T)\-/hh in the following. A second term
containing one hole and one particle contraction, (T)\-/ph. And finally, a
term which contains two particle contractions, (T)\-/pp. The diagrammatic
representations of these three terms is given in Fig.~\ref{fig:t3}(a)--(c) for
non-permuted terms.  All three terms and permutations of the particles and
holes add up to the full (T) energy.

We now analyze these three terms individually. In the following, we write the
algebraic contributions without the intermediates defined in
section~\ref{sec:triples}.

We start the analysis with the (T)\-/hh contribution, shown in
Fig.~\ref{fig:t3}(a), which reads without applying permutations
\begin{equation}
\label{eq:thhconv}
E = 8 \sum_{ijk} \sum_{abc} \sum_{mn} t_{in}^{ab} t_{im}^{ab}
  \upsilon_{jk}^{nc} \upsilon_{mc}^{jk} \frac{1}{\Delta^{ijk}_{abc}} \text.
\end{equation}
We first examine the BSIE of this contribution using MP2 amplitudes.
Fig.~\ref{fig:t3}(d) shows that the BSIE converges rapidly as
$N_\mathrm{v}^{-7/3}$.
The expression contains the sum over three virtual states. The sum over the
state $c$ does not provide an energy contribution for sufficiently high lying
virtual states. This is because the appearing Coulomb integrals contain three
states from the set of occupied orbitals.  Consequently, all Coulomb integrals
where the wave vector of the virtual state exceeds $3 k_\textrm{F}$ are zero
because of momentum conservation.  Therefore, an energy contribution from such
high lying virtual states can only stem from the sums over the states $a$ and
$b$. As mentioned earlier, contributions of type $t_{ij}^{\alpha b}$ have a
negligibly small contribution.  These contributions, using the notation of
augmented virtual states, can be written as
\begin{equation}
\label{eq:thhaug}
E = 8 \sum_{ijk} \sum_{\alpha \beta c} \sum_{mn}
  t_{in}^{\alpha \beta} t_{im}^{\alpha \beta}
  \upsilon_{jk}^{nc} \upsilon_{mc}^{jk}
  \frac{1}{\Delta^{ijk}_{\alpha \beta c}} \text.
\end{equation}
It is straightforward to derive the fundamental convergence behavior of
Eq.~(\ref{eq:thhaug}). In the limit of virtual states with high energy the wave vectors of
states inside the Fermi sphere become negligible and the amplitudes converge
as $q^{-4}$. The energy denominator $1/\Delta_{ijk}^{\alpha \beta c}$ converges
as $q^{-2}$. Thus, in the fundamental limit the energy expression can be
expressed as $\lim_{q \to \infty} \int_q^{\infty} dq' q'^2 \frac{1}{q'^4}
\frac{1}{q'^4} \frac{1}{q'^2}$, which is in accordance with the observed
$N_\mathrm{v}^{-7/3}$ behavior of the BSIE.  When applying the permutations in
Eqs.~(\ref{eq:corr_triples}) and (\ref{eq:perm2}) the BSIE of some
contributions is zero for sufficiently large transfer momenta $q$ as none of
the additional contributions fulfill the corresponding momentum conservation.

We now consider the case, where the CCD amplitudes are employed in the energy expression
for $E^\mathrm{(T)-hh}$.
The BSIE for this case is depicted in Fig.~\ref{fig:t3}(e) and
exhibits a clear $N_\mathrm{v}^{-1}$ behavior.
%
%
The identical effect was observed in sections~\ref{sec:otherLadder},~\ref{sec:linearRing},
and~\ref{sec:otherQuadratic}, and can be explained by
basis set incompleteness of the employed amplitudes $t_{ij}^{ab}$ within the
finite basis set.

%
%
The (T)\-/ph contribution, depicted in Fig.~\ref{fig:t3}(b), shows the same
convergence behavior as (T)\-/hh only when amplitudes from CCD are employed.
However, as shown in Fig.~\ref{fig:t3}(d), employing MP2 amplitudes the BSIE
converges with $N^{-5/3}$. Interestingly, the term without permutations, as
illustrated in Fig.~\ref{fig:t3}(b), converges faster, namely as $N^{-7/3}$.
Here, one of the Coulomb integrals contains three occupied states which prevents
any contributions from sufficiently high lying states $c$. This implies that
the momentum transfer of the other appearing Coulomb integral
$\upsilon_{bc}^{kf}$ is also restricted to a multiple of $k_\textrm{F}$.
However, when employing the permutations in Eqs.~(\ref{eq:corr_triples}) and
(\ref{eq:perm2}) some of the contributions show a slower convergence rate,
namely $N^{-5/3}$. This can be seen exemplarily for the the following
contribution
\begin{equation}
\label{eq:tphaug}
E = -4 \sum_{ijk} \sum_{\alpha \beta c} \sum_{me}
  t_{im}^{\alpha \beta} t_{ij}^{ce}
  \upsilon_{\beta \alpha}^{ek} \upsilon_{jk}^{mc}
  \frac{1}{\Delta^{ijk}_{\alpha \beta c}} \text.
\end{equation}
This energy expression is one particular contribution of the full (T) energy,
as given in Eq.~(\ref{eq:corr_triples}). It is obtained by taking only
contraction of the first element ($\overline{W}_{ijk}^{abc}$) in the
parenthesis of Eq.~(\ref{eq:corr_triples}), where $\overline{W}_{ijk}^{abc}$
does not contain all permutation from Eq.~(\ref{eq:perm2}) but only the one
where $a$ and $c$ are swapped.
In Eq.~(\ref{eq:tphaug}) the virtual- and augmented virtual states are chosen
such that they lead to the dominant contribution for this term.  In this
expression one of the occurring set of amplitudes contain states from the
augmented virtual state, whereas the other can incorporate states from the
finite virtual basis set. Consequently, the expression is piling up the factors
$q^{-4}$, $q^{-2}$, and $q^{-2}$ from the amplitudes, Coulomb integral and
energy denominator, respectively. The final BSIE is found by the corresponding
limit $\lim_{q \to \infty} \int_{q}^{\infty} dq' q'^2 \frac{1}{q'^4}
\frac{1}{q'^2} \frac{1}{q'^2}$.

%
%
Finally, we discuss the contribution (T)\-/pp, shown in Fig.~\ref{fig:t3}(c),
which shows a much larger BSIE than the two other discussed contributions,
as can be seen from Fig.~\ref{fig:t3}(d) and Fig.~\ref{fig:t3}(e).
The BSIE of this contribution is $N_\mathrm{v}^{-1}$ for both MP2 and CCD
amplitudes. Here both appearing Coulomb interactions contain three particle and
one hole state. Now, amplitudes containing states from the finite virtual basis
set can couple with augmented virtual states by the Coulomb interactions. The
corresponding expression, written with the notation of the finite and augmented
virtual states reads
\begin{equation}
\label{eq:tppaug}
E = 8 \sum_{ijk} \sum_{a \beta \gamma} \sum_{ef}
  t_{ij}^{af} t_{ij}^{ae}
  \upsilon_{\beta k}^{f \gamma } \upsilon_{e \gamma}^{\beta k}
  \frac{1}{\Delta^{ijk}_{a \beta \gamma}} \text{.}
\end{equation}
This leads to a completely different fundamental limit.
The two occurring amplitudes belong to the finite virtual basis set and do not
introduce a further power of $q^{-4}$ , however the Coulomb interactions now scale
as $q^{-2}$ resulting in a BSIE proportional to
$\lim_{q \to
\infty} \int_{q}^{\infty} dq' q'^2 \frac{1}{q'^2} \frac{1}{q'^2}
\frac{1}{q'^2}$. These terms, as given in Eq.~(\ref{eq:tppaug}), dominate the
BSIE of the (T)\-/pp term and therefore the BSIE of the whole (T) contribution.
Contributions where the amplitudes contain states from the augmented virtual
basis set show a convergence of $N^{-5/3}$ or faster.
We note that some of the permutations in Eqs.~(\ref{eq:corr_triples}) and
(\ref{eq:perm2}) show faster converging contributions.
We emphasize that these findings are not entirely novel. Empirically, it is
well\-/known that the leading rate of convergence of (T) is similar to
MP2~\cite{Barnes2008}. This even led to ad-hoc correction schemes that rescale
the (T) contributions with MP2 terms to correct for the BSIE of the
(T)~\cite{Knizia2009,Kallay2021}. Additionally, K{\"o}hn incorporated the
explicitly correlated framework for the (T) contribution~\cite{Koehn2009}. He was
able to numerically identify that the terms given in
Eq.~(\ref{eq:tppaug}) cover the important contributions to the
BSIE~\cite{Koehn2010}.

\subsection{(T) vs. (cT)}
\label{sec:cT}

\begin{figure*}[t]
\includegraphics{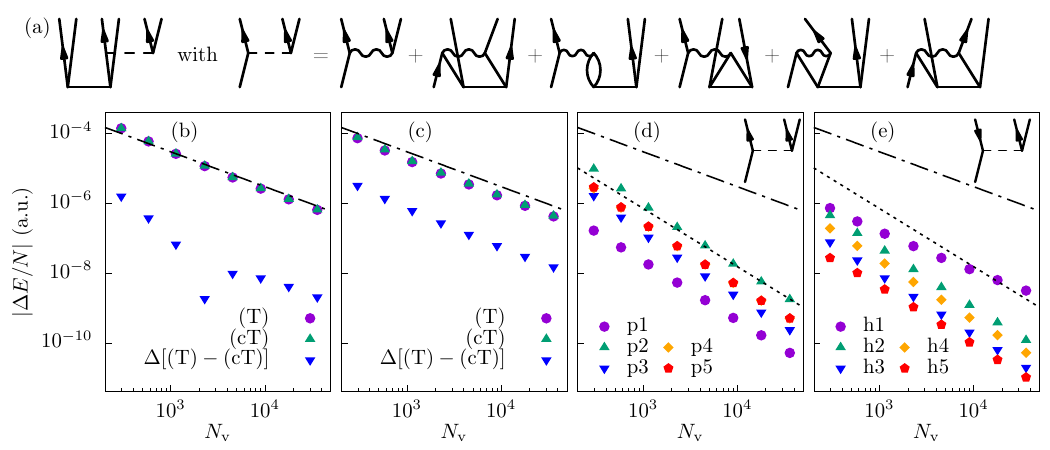}
\caption { (a) Diagrammatic illustration of the terms in (cT) theory as
given in Eq.~\ref{eq:Jp}. All terms connecting to another doubles amplitude on
the right are additional terms of (cT), labeled 1 to 5 in order of their
appearance. These terms are not occurring in the (T) theory.  (b--c) Shown is
the BSIE per electron of the (T) and (cT) contribution for a system with 14
electrons at a density $r_\mathrm{s}=5.0$~a.u.  Further shown is the
convergence of the energy difference between (T) and (cT).  (b),(c) show
results for MP2 and CCD, respectively. (d--e): p1--5 are the five additional
diagrams emerging from Eq.~(\ref{eq:Jp}).  h1--5 are the corresponding terms
connected from Eq~(\ref{eq:Jh}). CBS estimates are obtained from extrapolation
of the two largest
systems using the corresponding power law.
}
 \label{fig:sf_t}
\end{figure*}


In this section, we analyze the BSIE of the recently proposed complete
perturbative triples correction (cT). The motivation for this work was to
incorporate additional contributions to prevent the perturbative correction
from diverging for metallic systems~\cite{Masios2023} in the thermodynamic limit.
These further
contributions are given in Eqs.~(\ref{eq:Jp}) and~(\ref{eq:Jh}). In total ten
further terms beyond the bare Coulomb interactions in Eqs.~(\ref{eq:Jp}) and~(\ref{eq:Jh})
are included. In passing, we note that only two of these additional
terms are accountable for resolving the divergence for metallic systems in the
(T) method, namely the two ring terms
$\upsilon_{ef}^{bm} t_{km}^{cf}$ and $\upsilon_{jf}^{mn} t_{kn}^{cf}$.
The performed analysis in the previous work focused on the
small\-/$q$ regime where the divergence occurs. In this work, we analyze the
behavior for large $q$, i.e.\ studying the BSIE of the (cT) model.

For these reasons, we analyze the convergence of the energy expressions (T) and
(cT) for the same system. Similar to the analysis in CCD, the BSIE is shown for
MP2 amplitudes and converged CCD amplitudes, shown in Fig.~\ref{fig:sf_t}(b)
and (c), respectively. It can be seen that the BSIE of the additional terms in
(cT) are at least one order of magnitude smaller than the BSIE of the (T)
expression. The difference between (T) and (cT) is larger when using CCD
amplitudes, however, this difference shows also a slow $N_\mathrm{v}^{-1}$
convergence when using MP2 amplitudes.  This analysis reveals that at least one
of the additional terms is expected to cause a slow $N_\mathrm{v}^{-1}$
convergence.  Consequently, we perform an analysis of the ten additional
contributions individually.  The diagrammatic contributions are depicted in
Fig.~\ref{fig:sf_t}(a). We note that the first diagram on the right\-/hand side
corresponds to (T), whereas all others are included in (cT) and are referred to
as h/p1--h/p5.  Again, we use amplitudes from MP2 level of theory for the
following analysis as all terms show an $N_\mathrm{v}^{-1}$ with CCD amplitudes.

The BSIEs of the individual terms are given in
Fig.~\ref{fig:sf_t}(d) and (e).  Evidently, only one of these terms is slowly
converging, whereas all other contributions converge as $N_\mathrm{v}^{-5/3}$.
Interestingly, the slowly converging h1 contribution is very
similar to one of the slowly convergent terms in the CCD expression (viz.\
Fig.~\ref{fig:energy_channels}(d) ).

\subsection{Dependence on electron number and density}
\label{sec:dep_num_tri}

\begin{widetext}

\begin{figure}
\includegraphics{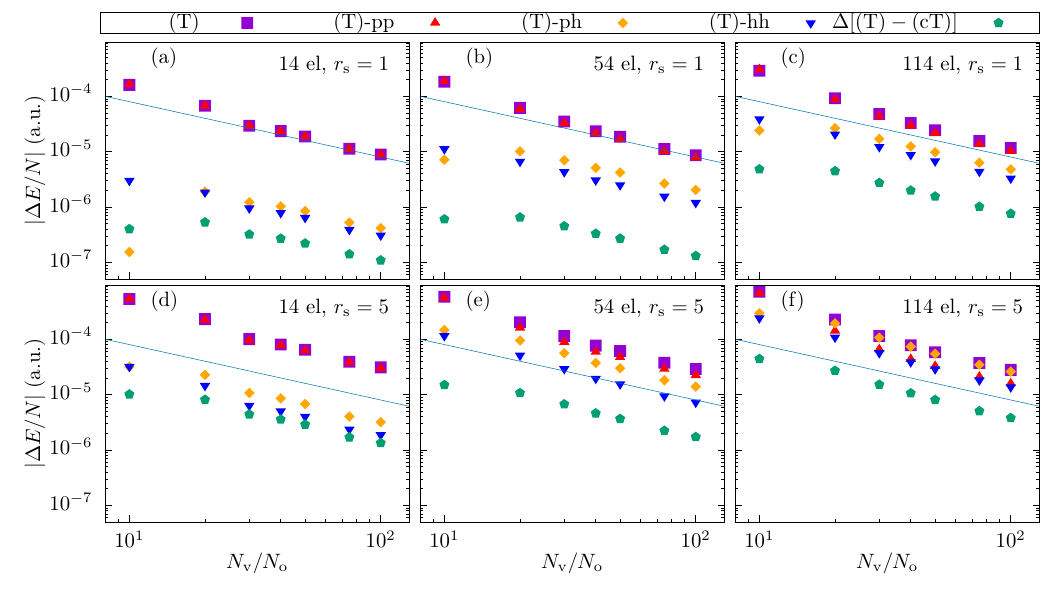}
\caption {
BSIE per electron for different contributions are shown.  (a)--(c) show results
for $r_\mathrm{s}=1$, whereas (d)--(f) uses a density of $r_\mathrm{s}=5$~a.u.
Panels in the left, middle, and right column show different system sizes with
14, 54, and 114 electrons, respectively. To allow comparison between systems
with different electron numbers we present the results in BSIE per electron,
and the number of virtual states per occupied orbitals. The blue line is
proportional to $N^{-1}$ and allows comparison between the six different
panels.
}
\label{fig:el_number_t}
\end{figure}
\end{widetext}

Up to here the analysis of the BSIE of the triples contributions was done for a
system with 14 electrons and a density of $r_\mathrm{s}=5$~a.u..  In this section,
we want to study two different densities, $r_\mathrm{s}=1$~a.u. and $r_\mathrm{s}=5$~a.u.,
as well as three different system sizes, namely 14, 54, and 114 electrons.
Fig.~\ref{fig:el_number_t} summarizes the BSIE of the quantities discussed in
the previous sections, viz. the contributions of the previously defined
channels T-pp, T-ph, and T-hh. And further the difference between the (T) and
the (cT) method.

For all analyzed systems, the difference between (T) and (cT) is at least
around one order of magnitude  smaller than the BSIE of (T). It can be observed
that the terms beyond (T) in (cT) are more important for larger electron
numbers as well as for larger $r_\mathrm{s}$.

Moving on to the discussion of the three channels (T)-pp, (T)-ph, and (T)-hh.
For small values of $r_\mathrm{s}$, as well as for small electron numbers,
the (T)-pp dominates the BSIE of the overall BSIE of (T).  However, the BSIE of
the other two channels increases on a relative scale if the number of
electrons or the value of $r_\mathrm{s}$ is increased, respectively.  For the
system with 114 electrons and a density of $r_\mathrm{s}=5.0$~a.u. the (T)-ph term
is already the largest in magnitude.  It is important to note that, as
discussed in the previous section, the $N_\mathrm{v}^{-1}$ decay of (T)-ph and
(T)-hh does not stem from contributions with large momenta $q$.  The BSIE
rather originates from the changes in the CCD amplitude elements corresponding
to the finite virtual basis set. These two leading order contributions, namely
the basis set incompleteness in the CCD amplitudes and the additional
contributions from $W_{ijk}^{a \beta \gamma}$ as given in Eq.~(\ref{eq:tppaug})
have been already identified by K{\"o}hn~\cite{Koehn2009,Koehn2010} in the
context of an F12 correction for the (T) contribution. The analysis performed
here shows that the relative importance of one or the other contribution
depends strongly on the studied system.

\section{Summary and conclusions}

In this work, we present a detailed analysis of different coupled\-/cluster
theories applied to the UEG in the large-momentum-transfer limit.
This allows a
fundamental investigation of the BSIE in these theories.  Even though it is
well-known that the MP2 and particle--particle ladder contributions are the most dominant terms in
CCD, this work investigates all further terms in CCD.
In particular, we have shown that in MP3 theory, which corresponds to a low-order approximation
of CCD (here referred to as the linear terms of CCD$^{(1)}$),
all diagrammatic contributions besides MP2 and PPL converge
fundamentally faster, namely as $N_\mathrm{v}^{-5/3}$ or $N_\mathrm{v}^{-7/3}$.
In CCD theory, however, all diagrammatic contributions converge
slowly as $N_\mathrm{v}^{-1}$. This is due to the fact that the PPL term
and other contributions couple amplitude elements with
small and large momenta.  Nevertheless, it has been shown for different
densities and electron numbers, that MP2 and PPL are throughout the dominant
terms.  This is also evident in the transition structure factor, which shows
only for the MP2 and PPL term a $q^{-4}$ decay for large values of $q$. All
other diagrammatic contributions show a faster convergence behavior.  It was
further shown that these key findings are valid in the range of realistic
densities, $1.0 \le r_\mathrm{s} \le 5.0$ and show only little dependence on
the number of electrons in the system.

For the (T) contribution a similar decomposition into three channels
was carried out. It was shown that only one of the
channels shows a leading order contribution when using MP2 amplitudes.
In CCD, however, due to the basis
set incompleteness of the employed doubles amplitudes, all three channels show
the same $N_\textrm{v}^{-1}$ decay. The relative strength of the three
different channels to the total BSIE depends on the number of electrons and the
employed density. It was shown that there are essentially two major contribution to
the overall BSIE: (i) a term which stems from elements with large momentum
transfer on the additional Coulomb interaction of (T),
which is only dominantly present in the so-called (T)-pp channel and (ii)
contributions which are linked to the basis set
incompleteness of the underlying CCD amplitudes. The latter contribution is
significant for all three different channels, (T)-pp, (T)-ph, and (T)-hh.

Finally, the BSIE of the recently proposed (cT) approach was investigated. It
was shown that the additional contributions beyond the (T) approach show only a
small BSIE. This is desirable as this new approach was conceived as a
non-diverging perturbative triples correction for vanishing-gap systems. The
mentioned divergence occurs for small momentum transfers,
whereas the BSIE is attributed
mainly in the regime of large momentum transfers.

\section{Acknowledgments}
The authors thankfully acknowledge support and funding from the European
Research Council (ERC) under the European Union’s Horizon 2020 research and
innovation program (Grant Agreement No. 715594). We gratefully acknowledge many
fruitful discussions with Alejandro Gallo.  The computational results presented
have been achieved in part using the Vienna Scientific Cluster (VSC).

\bibliography{uegcbs}

\begin{thebibliography}{59}%
\makeatletter
\providecommand \@ifxundefined [1]{%
 \@ifx{#1\undefined}
}%
\providecommand \@ifnum [1]{%
 \ifnum #1\expandafter \@firstoftwo
 \else \expandafter \@secondoftwo
 \fi
}%
\providecommand \@ifx [1]{%
 \ifx #1\expandafter \@firstoftwo
 \else \expandafter \@secondoftwo
 \fi
}%
\providecommand \natexlab [1]{#1}%
\providecommand \enquote  [1]{``#1''}%
\providecommand \bibnamefont  [1]{#1}%
\providecommand \bibfnamefont [1]{#1}%
\providecommand \citenamefont [1]{#1}%
\providecommand \href@noop [0]{\@secondoftwo}%
\providecommand \href [0]{\begingroup \@sanitize@url \@href}%
\providecommand \@href[1]{\@@startlink{#1}\@@href}%
\providecommand \@@href[1]{\endgroup#1\@@endlink}%
\providecommand \@sanitize@url [0]{\catcode `\\12\catcode `\$12\catcode
  `\&12\catcode `\#12\catcode `\^12\catcode `\_12\catcode `\%12\relax}%
\providecommand \@@startlink[1]{}%
\providecommand \@@endlink[0]{}%
\providecommand \url  [0]{\begingroup\@sanitize@url \@url }%
\providecommand \@url [1]{\endgroup\@href {#1}{\urlprefix }}%
\providecommand \urlprefix  [0]{URL }%
\providecommand \Eprint [0]{\href }%
\providecommand \doibase [0]{https://doi.org/}%
\providecommand \selectlanguage [0]{\@gobble}%
\providecommand \bibinfo  [0]{\@secondoftwo}%
\providecommand \bibfield  [0]{\@secondoftwo}%
\providecommand \translation [1]{[#1]}%
\providecommand \BibitemOpen [0]{}%
\providecommand \bibitemStop [0]{}%
\providecommand \bibitemNoStop [0]{.\EOS\space}%
\providecommand \EOS [0]{\spacefactor3000\relax}%
\providecommand \BibitemShut  [1]{\csname bibitem#1\endcsname}%
\let\auto@bib@innerbib\@empty
\bibitem [{\citenamefont {Bartlett}\ and\ \citenamefont
  {Musia\l{}}(2007)}]{bartlett2007}%
  \BibitemOpen
  \bibfield  {author} {\bibinfo {author} {\bibfnamefont {R.~J.}\ \bibnamefont
  {Bartlett}}\ and\ \bibinfo {author} {\bibfnamefont {M.}~\bibnamefont
  {Musia\l{}}},\ }\bibfield  {title} {\bibinfo {title} {Coupled-cluster theory
  in quantum chemistry},\ }\href {https://doi.org/10.1103/RevModPhys.79.291}
  {\bibfield  {journal} {\bibinfo  {journal} {Rev. Mod. Phys.}\ }\textbf
  {\bibinfo {volume} {79}},\ \bibinfo {pages} {291} (\bibinfo {year}
  {2007})}\BibitemShut {NoStop}%
\bibitem [{\citenamefont {Voloshina}\ \emph {et~al.}(2011)\citenamefont
  {Voloshina}, \citenamefont {Usvyat}, \citenamefont {Sch\"utz}, \citenamefont
  {Dedkov},\ and\ \citenamefont {Paulus}}]{Voloshina2011}%
  \BibitemOpen
  \bibfield  {author} {\bibinfo {author} {\bibfnamefont {E.}~\bibnamefont
  {Voloshina}}, \bibinfo {author} {\bibfnamefont {D.}~\bibnamefont {Usvyat}},
  \bibinfo {author} {\bibfnamefont {M.}~\bibnamefont {Sch\"utz}}, \bibinfo
  {author} {\bibfnamefont {Y.}~\bibnamefont {Dedkov}},\ and\ \bibinfo {author}
  {\bibfnamefont {B.}~\bibnamefont {Paulus}},\ }\bibfield  {title} {\bibinfo
  {title} {On the physisorption of water on graphene: a ccsd(t) study},\ }\href
  {https://doi.org/10.1039/C1CP20609E} {\bibfield  {journal} {\bibinfo
  {journal} {Phys. Chem. Chem. Phys.}\ }\textbf {\bibinfo {volume} {13}},\
  \bibinfo {pages} {12041} (\bibinfo {year} {2011})}\BibitemShut {NoStop}%
\bibitem [{\citenamefont {Kubas}\ \emph {et~al.}(2016)\citenamefont {Kubas},
  \citenamefont {Berger}, \citenamefont {Oberhofer}, \citenamefont {Maganas},
  \citenamefont {Reuter},\ and\ \citenamefont {Neese}}]{kubas2016}%
  \BibitemOpen
  \bibfield  {author} {\bibinfo {author} {\bibfnamefont {A.}~\bibnamefont
  {Kubas}}, \bibinfo {author} {\bibfnamefont {D.}~\bibnamefont {Berger}},
  \bibinfo {author} {\bibfnamefont {H.}~\bibnamefont {Oberhofer}}, \bibinfo
  {author} {\bibfnamefont {D.}~\bibnamefont {Maganas}}, \bibinfo {author}
  {\bibfnamefont {K.}~\bibnamefont {Reuter}},\ and\ \bibinfo {author}
  {\bibfnamefont {F.}~\bibnamefont {Neese}},\ }\bibfield  {title} {\bibinfo
  {title} {Surface adsorption energetics studied with “gold standard”
  wave-function-based ab initio methods: Small-molecule binding to tio2(110)},\
  }\href {https://doi.org/10.1021/acs.jpclett.6b01845} {\bibfield  {journal}
  {\bibinfo  {journal} {J. Phys. Chem. Lett}\ }\textbf {\bibinfo {volume}
  {7}},\ \bibinfo {pages} {4207} (\bibinfo {year} {2016})}\BibitemShut
  {NoStop}%
\bibitem [{\citenamefont {Brandenburg}\ \emph {et~al.}(2019)\citenamefont
  {Brandenburg}, \citenamefont {Zen}, \citenamefont {Fitzner}, \citenamefont
  {Ramberger}, \citenamefont {Kresse}, \citenamefont {Tsatsoulis},
  \citenamefont {Grüneis}, \citenamefont {Michaelides},\ and\ \citenamefont
  {Alfè}}]{Brandenburg2019}%
  \BibitemOpen
  \bibfield  {author} {\bibinfo {author} {\bibfnamefont {J.~G.}\ \bibnamefont
  {Brandenburg}}, \bibinfo {author} {\bibfnamefont {A.}~\bibnamefont {Zen}},
  \bibinfo {author} {\bibfnamefont {M.}~\bibnamefont {Fitzner}}, \bibinfo
  {author} {\bibfnamefont {B.}~\bibnamefont {Ramberger}}, \bibinfo {author}
  {\bibfnamefont {G.}~\bibnamefont {Kresse}}, \bibinfo {author} {\bibfnamefont
  {T.}~\bibnamefont {Tsatsoulis}}, \bibinfo {author} {\bibfnamefont
  {A.}~\bibnamefont {Grüneis}}, \bibinfo {author} {\bibfnamefont
  {A.}~\bibnamefont {Michaelides}},\ and\ \bibinfo {author} {\bibfnamefont
  {D.}~\bibnamefont {Alfè}},\ }\bibfield  {title} {\bibinfo {title}
  {Physisorption of water on graphene: Subchemical accuracy from many-body
  electronic structure methods},\ }\href
  {https://doi.org/10.1021/acs.jpclett.8b03679} {\bibfield  {journal} {\bibinfo
   {journal} {J. Phys. Chem. Lett.}\ }\textbf {\bibinfo {volume} {10}},\
  \bibinfo {pages} {358} (\bibinfo {year} {2019})}\BibitemShut {NoStop}%
\bibitem [{\citenamefont {Tsatsoulis}\ \emph {et~al.}(2018)\citenamefont
  {Tsatsoulis}, \citenamefont {Sakong}, \citenamefont {Groß},\ and\
  \citenamefont {Grüneis}}]{tsatsoulis2018}%
  \BibitemOpen
  \bibfield  {author} {\bibinfo {author} {\bibfnamefont {T.}~\bibnamefont
  {Tsatsoulis}}, \bibinfo {author} {\bibfnamefont {S.}~\bibnamefont {Sakong}},
  \bibinfo {author} {\bibfnamefont {A.}~\bibnamefont {Groß}},\ and\ \bibinfo
  {author} {\bibfnamefont {A.}~\bibnamefont {Grüneis}},\ }\bibfield  {title}
  {\bibinfo {title} {Reaction energetics of hydrogen on si(100) surface: A
  periodic many-electron theory study},\ }\href
  {https://doi.org/10.1063/1.5055706} {\bibfield  {journal} {\bibinfo
  {journal} {J. Chem. Phys.}\ }\textbf {\bibinfo {volume} {149}},\ \bibinfo
  {pages} {244105} (\bibinfo {year} {2018})}\BibitemShut {NoStop}%
\bibitem [{\citenamefont {Sauer}(2019)}]{Sauer2019}%
  \BibitemOpen
  \bibfield  {author} {\bibinfo {author} {\bibfnamefont {J.}~\bibnamefont
  {Sauer}},\ }\bibfield  {title} {\bibinfo {title} {Ab initio calculations for
  molecule–surface interactions with chemical accuracy},\ }\href
  {https://doi.org/10.1021/acs.accounts.9b00506} {\bibfield  {journal}
  {\bibinfo  {journal} {Acc. Chem. Res.}\ }\textbf {\bibinfo {volume} {52}},\
  \bibinfo {pages} {3502} (\bibinfo {year} {2019})}\BibitemShut {NoStop}%
\bibitem [{\citenamefont {Lau}\ \emph {et~al.}(2021)\citenamefont {Lau},
  \citenamefont {Knizia},\ and\ \citenamefont {Berkelbach}}]{Lau2021}%
  \BibitemOpen
  \bibfield  {author} {\bibinfo {author} {\bibfnamefont {B.~T.~G.}\
  \bibnamefont {Lau}}, \bibinfo {author} {\bibfnamefont {G.}~\bibnamefont
  {Knizia}},\ and\ \bibinfo {author} {\bibfnamefont {T.~C.}\ \bibnamefont
  {Berkelbach}},\ }\bibfield  {title} {\bibinfo {title} {Regional embedding
  enables high-level quantum chemistry for surface science},\ }\href
  {https://doi.org/10.1021/acs.jpclett.0c03274} {\bibfield  {journal} {\bibinfo
   {journal} {J. Phys. Chem. Lett.}\ }\textbf {\bibinfo {volume} {12}},\
  \bibinfo {pages} {1104} (\bibinfo {year} {2021})}\BibitemShut {NoStop}%
\bibitem [{\citenamefont {Schäfer}\ \emph {et~al.}(2021)\citenamefont
  {Schäfer}, \citenamefont {Gallo}, \citenamefont {Irmler}, \citenamefont
  {Hummel},\ and\ \citenamefont {Grüneis}}]{Schaefer2021}%
  \BibitemOpen
  \bibfield  {author} {\bibinfo {author} {\bibfnamefont {T.}~\bibnamefont
  {Schäfer}}, \bibinfo {author} {\bibfnamefont {A.}~\bibnamefont {Gallo}},
  \bibinfo {author} {\bibfnamefont {A.}~\bibnamefont {Irmler}}, \bibinfo
  {author} {\bibfnamefont {F.}~\bibnamefont {Hummel}},\ and\ \bibinfo {author}
  {\bibfnamefont {A.}~\bibnamefont {Grüneis}},\ }\bibfield  {title} {\bibinfo
  {title} {Surface science using coupled cluster theory via local wannier
  functions and in-rpa-embedding: The case of water on graphitic carbon
  nitride},\ }\href {https://doi.org/10.1063/5.0074936} {\bibfield  {journal}
  {\bibinfo  {journal} {J. Chem. Phys.}\ }\textbf {\bibinfo {volume} {155}},\
  \bibinfo {pages} {244103} (\bibinfo {year} {2021})}\BibitemShut {NoStop}%
\bibitem [{\citenamefont {Mullan}\ \emph {et~al.}(2022)\citenamefont {Mullan},
  \citenamefont {Maschio}, \citenamefont {Saalfrank},\ and\ \citenamefont
  {Usvyat}}]{Mullan2022}%
  \BibitemOpen
  \bibfield  {author} {\bibinfo {author} {\bibfnamefont {T.}~\bibnamefont
  {Mullan}}, \bibinfo {author} {\bibfnamefont {L.}~\bibnamefont {Maschio}},
  \bibinfo {author} {\bibfnamefont {P.}~\bibnamefont {Saalfrank}},\ and\
  \bibinfo {author} {\bibfnamefont {D.}~\bibnamefont {Usvyat}},\ }\bibfield
  {title} {\bibinfo {title} {Reaction barriers on non-conducting surfaces
  beyond periodic local mp2: Diffusion of hydrogen on alpha-al2o3(0001) as a
  test case},\ }\href {https://doi.org/10.1063/5.0082805} {\bibfield  {journal}
  {\bibinfo  {journal} {J. Chem. Phys.}\ }\textbf {\bibinfo {volume} {156}},\
  \bibinfo {pages} {074109} (\bibinfo {year} {2022})}\BibitemShut {NoStop}%
\bibitem [{\citenamefont {Shi}\ \emph {et~al.}(2023)\citenamefont {Shi},
  \citenamefont {Zen}, \citenamefont {Kapil}, \citenamefont {Nagy},
  \citenamefont {Grüneis},\ and\ \citenamefont {Michaelides}}]{Shi2023}%
  \BibitemOpen
  \bibfield  {author} {\bibinfo {author} {\bibfnamefont {B.~X.}\ \bibnamefont
  {Shi}}, \bibinfo {author} {\bibfnamefont {A.}~\bibnamefont {Zen}}, \bibinfo
  {author} {\bibfnamefont {V.}~\bibnamefont {Kapil}}, \bibinfo {author}
  {\bibfnamefont {P.~R.}\ \bibnamefont {Nagy}}, \bibinfo {author}
  {\bibfnamefont {A.}~\bibnamefont {Grüneis}},\ and\ \bibinfo {author}
  {\bibfnamefont {A.}~\bibnamefont {Michaelides}},\ }\bibfield  {title}
  {\bibinfo {title} {Many-body methods for surface chemistry come of age:
  Achieving consensus with experiments},\ }\href
  {https://doi.org/10.1021/jacs.3c09616} {\bibfield  {journal} {\bibinfo
  {journal} {J. Am. Chem. Soc.}\ }\textbf {\bibinfo {volume} {145}},\ \bibinfo
  {pages} {25372} (\bibinfo {year} {2023})}\BibitemShut {NoStop}%
\bibitem [{\citenamefont {Shepherd}\ and\ \citenamefont
  {Gr\"uneis}(2013)}]{Shepherd2013}%
  \BibitemOpen
  \bibfield  {author} {\bibinfo {author} {\bibfnamefont {J.~J.}\ \bibnamefont
  {Shepherd}}\ and\ \bibinfo {author} {\bibfnamefont {A.}~\bibnamefont
  {Gr\"uneis}},\ }\bibfield  {title} {\bibinfo {title} {Many-body quantum
  chemistry for the electron gas: Convergent perturbative theories},\ }\href
  {https://doi.org/10.1103/PhysRevLett.110.226401} {\bibfield  {journal}
  {\bibinfo  {journal} {Phys. Rev. Lett.}\ }\textbf {\bibinfo {volume} {110}},\
  \bibinfo {pages} {226401} (\bibinfo {year} {2013})}\BibitemShut {NoStop}%
\bibitem [{\citenamefont {Masios}\ \emph {et~al.}(2023)\citenamefont {Masios},
  \citenamefont {Irmler}, \citenamefont {Sch\"afer},\ and\ \citenamefont
  {Gr\"uneis}}]{Masios2023}%
  \BibitemOpen
  \bibfield  {author} {\bibinfo {author} {\bibfnamefont {N.}~\bibnamefont
  {Masios}}, \bibinfo {author} {\bibfnamefont {A.}~\bibnamefont {Irmler}},
  \bibinfo {author} {\bibfnamefont {T.}~\bibnamefont {Sch\"afer}},\ and\
  \bibinfo {author} {\bibfnamefont {A.}~\bibnamefont {Gr\"uneis}},\ }\bibfield
  {title} {\bibinfo {title} {Averting the infrared catastrophe in the gold
  standard of quantum chemistry},\ }\href
  {https://doi.org/10.1103/PhysRevLett.131.186401} {\bibfield  {journal}
  {\bibinfo  {journal} {Phys. Rev. Lett.}\ }\textbf {\bibinfo {volume} {131}},\
  \bibinfo {pages} {186401} (\bibinfo {year} {2023})}\BibitemShut {NoStop}%
\bibitem [{\citenamefont {Shepherd}\ \emph {et~al.}(2012)\citenamefont
  {Shepherd}, \citenamefont {Gr{\"{u}}neis}, \citenamefont {Booth},
  \citenamefont {Kresse},\ and\ \citenamefont {Alavi}}]{shepherd2012}%
  \BibitemOpen
  \bibfield  {author} {\bibinfo {author} {\bibfnamefont {J.~J.}\ \bibnamefont
  {Shepherd}}, \bibinfo {author} {\bibfnamefont {A.}~\bibnamefont
  {Gr{\"{u}}neis}}, \bibinfo {author} {\bibfnamefont {G.~H.}\ \bibnamefont
  {Booth}}, \bibinfo {author} {\bibfnamefont {G.}~\bibnamefont {Kresse}},\ and\
  \bibinfo {author} {\bibfnamefont {A.}~\bibnamefont {Alavi}},\ }\bibfield
  {title} {\bibinfo {title} {Convergence of many-body wave-function expansions
  using a plane-wave basis: From homogeneous electron gas to solid state
  systems},\ }\href {https://doi.org/10.1103/PhysRevB.86.035111} {\bibfield
  {journal} {\bibinfo  {journal} {Phys. Rev. B}\ }\textbf {\bibinfo {volume}
  {86}},\ \bibinfo {pages} {035111} (\bibinfo {year} {2012})}\BibitemShut
  {NoStop}%
\bibitem [{\citenamefont {Gr\"{u}neis}\ \emph {et~al.}(2013)\citenamefont
  {Gr\"{u}neis}, \citenamefont {Shepherd}, \citenamefont {Alavi}, \citenamefont
  {Tew},\ and\ \citenamefont {Booth}}]{Grueneis2013}%
  \BibitemOpen
  \bibfield  {author} {\bibinfo {author} {\bibfnamefont {A.}~\bibnamefont
  {Gr\"{u}neis}}, \bibinfo {author} {\bibfnamefont {J.~J.}\ \bibnamefont
  {Shepherd}}, \bibinfo {author} {\bibfnamefont {A.}~\bibnamefont {Alavi}},
  \bibinfo {author} {\bibfnamefont {D.~P.}\ \bibnamefont {Tew}},\ and\ \bibinfo
  {author} {\bibfnamefont {G.~H.}\ \bibnamefont {Booth}},\ }\bibfield  {title}
  {\bibinfo {title} {Explicitly correlated plane waves: Accelerating
  convergence in periodic wavefunction expansions},\ }\href
  {https://doi.org/10.1063/1.4818753} {\bibfield  {journal} {\bibinfo
  {journal} {J. Chem. Phys.}\ }\textbf {\bibinfo {volume} {139}},\ \bibinfo
  {pages} {084112} (\bibinfo {year} {2013})}\BibitemShut {NoStop}%
\bibitem [{\citenamefont {Callahan}\ \emph {et~al.}(2021)\citenamefont
  {Callahan}, \citenamefont {Lange},\ and\ \citenamefont
  {Berkelbach}}]{Callahan2021}%
  \BibitemOpen
  \bibfield  {author} {\bibinfo {author} {\bibfnamefont {J.~M.}\ \bibnamefont
  {Callahan}}, \bibinfo {author} {\bibfnamefont {M.~F.}\ \bibnamefont
  {Lange}},\ and\ \bibinfo {author} {\bibfnamefont {T.~C.}\ \bibnamefont
  {Berkelbach}},\ }\bibfield  {title} {\bibinfo {title} {{Dynamical correlation
  energy of metals in large basis sets from downfolding and composite
  approaches}},\ }\href {https://doi.org/10.1063/5.0049890} {\bibfield
  {journal} {\bibinfo  {journal} {J. Chem. Phys.}\ }\textbf {\bibinfo {volume}
  {154}},\ \bibinfo {pages} {211105} (\bibinfo {year} {2021})}\BibitemShut
  {NoStop}%
\bibitem [{\citenamefont {Kato}(1957)}]{kato_1957}%
  \BibitemOpen
  \bibfield  {author} {\bibinfo {author} {\bibfnamefont {T.}~\bibnamefont
  {Kato}},\ }\bibfield  {title} {\bibinfo {title} {On the eigenfunctions of
  many-particle systems in quantum mechanics},\ }\href
  {https://doi.org/10.1002/cpa.3160100201} {\bibfield  {journal} {\bibinfo
  {journal} {Commun. Pure Appl. Math.}\ }\textbf {\bibinfo {volume} {10}},\
  \bibinfo {pages} {151} (\bibinfo {year} {1957})}\BibitemShut {NoStop}%
\bibitem [{\citenamefont {Pack}\ and\ \citenamefont {Brown}(1966)}]{Pack1966}%
  \BibitemOpen
  \bibfield  {author} {\bibinfo {author} {\bibfnamefont {R.~T.}\ \bibnamefont
  {Pack}}\ and\ \bibinfo {author} {\bibfnamefont {W.~B.}\ \bibnamefont
  {Brown}},\ }\bibfield  {title} {\bibinfo {title} {{Cusp Conditions for
  Molecular Wavefunctions}},\ }\href {https://doi.org/10.1063/1.1727605}
  {\bibfield  {journal} {\bibinfo  {journal} {J. Chem. Phys.}\ }\textbf
  {\bibinfo {volume} {45}},\ \bibinfo {pages} {556} (\bibinfo {year}
  {1966})}\BibitemShut {NoStop}%
\bibitem [{\citenamefont {Morgan}\ and\ \citenamefont
  {Kutzelnigg}(1993)}]{Morgan1993}%
  \BibitemOpen
  \bibfield  {author} {\bibinfo {author} {\bibfnamefont {J.~D.}\ \bibnamefont
  {Morgan}}\ and\ \bibinfo {author} {\bibfnamefont {W.}~\bibnamefont
  {Kutzelnigg}},\ }\bibfield  {title} {\bibinfo {title} {Hund's rules, the
  alternating rule, and symmetry holes},\ }\href
  {https://doi.org/10.1021/j100112a051} {\bibfield  {journal} {\bibinfo
  {journal} {J. Phys. Chem.}\ }\textbf {\bibinfo {volume} {97}},\ \bibinfo
  {pages} {2425} (\bibinfo {year} {1993})}\BibitemShut {NoStop}%
\bibitem [{\citenamefont {Kutzelnigg}\ and\ \citenamefont
  {Klopper}(1991)}]{kutzelnigg_1991}%
  \BibitemOpen
  \bibfield  {author} {\bibinfo {author} {\bibfnamefont {W.}~\bibnamefont
  {Kutzelnigg}}\ and\ \bibinfo {author} {\bibfnamefont {W.}~\bibnamefont
  {Klopper}},\ }\bibfield  {title} {\bibinfo {title} {Wave functions with terms
  linear in the interelectronic coordinates to take care of the correlation
  cusp. i. general theory},\ }\href {https://doi.org/10.1063/1.459921}
  {\bibfield  {journal} {\bibinfo  {journal} {J. Chem. Phys.}\ }\textbf
  {\bibinfo {volume} {94}},\ \bibinfo {pages} {1985} (\bibinfo {year}
  {1991})}\BibitemShut {NoStop}%
\bibitem [{\citenamefont {Ten-no}(2007)}]{Tenno2007}%
  \BibitemOpen
  \bibfield  {author} {\bibinfo {author} {\bibfnamefont {S.}~\bibnamefont
  {Ten-no}},\ }\bibfield  {title} {\bibinfo {title} {{New implementation of
  second-order Møller-Plesset perturbation theory with an analytic Slater-type
  geminal}},\ }\href {https://doi.org/10.1063/1.2403853} {\bibfield  {journal}
  {\bibinfo  {journal} {J. Chem. Phys.}\ }\textbf {\bibinfo {volume} {126}},\
  \bibinfo {pages} {014108} (\bibinfo {year} {2007})}\BibitemShut {NoStop}%
\bibitem [{\citenamefont {Hättig}\ \emph {et~al.}(2012)\citenamefont
  {Hättig}, \citenamefont {Klopper}, \citenamefont {Köhn},\ and\
  \citenamefont {Tew}}]{Haettig2012}%
  \BibitemOpen
  \bibfield  {author} {\bibinfo {author} {\bibfnamefont {C.}~\bibnamefont
  {Hättig}}, \bibinfo {author} {\bibfnamefont {W.}~\bibnamefont {Klopper}},
  \bibinfo {author} {\bibfnamefont {A.}~\bibnamefont {Köhn}},\ and\ \bibinfo
  {author} {\bibfnamefont {D.~P.}\ \bibnamefont {Tew}},\ }\bibfield  {title}
  {\bibinfo {title} {Explicitly correlated electrons in molecules},\ }\href
  {https://doi.org/10.1021/cr200168z} {\bibfield  {journal} {\bibinfo
  {journal} {Chem. Rev.}\ }\textbf {\bibinfo {volume} {112}},\ \bibinfo {pages}
  {4} (\bibinfo {year} {2012})}\BibitemShut {NoStop}%
\bibitem [{\citenamefont {Kong}\ \emph {et~al.}(2012)\citenamefont {Kong},
  \citenamefont {Bischoff},\ and\ \citenamefont {Valeev}}]{Kong2012}%
  \BibitemOpen
  \bibfield  {author} {\bibinfo {author} {\bibfnamefont {L.}~\bibnamefont
  {Kong}}, \bibinfo {author} {\bibfnamefont {F.~A.}\ \bibnamefont {Bischoff}},\
  and\ \bibinfo {author} {\bibfnamefont {E.~F.}\ \bibnamefont {Valeev}},\
  }\bibfield  {title} {\bibinfo {title} {Explicitly correlated r12/f12 methods
  for electronic structure},\ }\href {https://doi.org/10.1021/cr200204r}
  {\bibfield  {journal} {\bibinfo  {journal} {Chem. Rev.}\ }\textbf {\bibinfo
  {volume} {112}},\ \bibinfo {pages} {75} (\bibinfo {year} {2012})}\BibitemShut
  {NoStop}%
\bibitem [{\citenamefont {Grüneis}\ \emph {et~al.}(2017)\citenamefont
  {Grüneis}, \citenamefont {Hirata}, \citenamefont {Ohnishi},\ and\
  \citenamefont {Ten-no}}]{Gruneis2017}%
  \BibitemOpen
  \bibfield  {author} {\bibinfo {author} {\bibfnamefont {A.}~\bibnamefont
  {Grüneis}}, \bibinfo {author} {\bibfnamefont {S.}~\bibnamefont {Hirata}},
  \bibinfo {author} {\bibfnamefont {Y.-y.}\ \bibnamefont {Ohnishi}},\ and\
  \bibinfo {author} {\bibfnamefont {S.}~\bibnamefont {Ten-no}},\ }\bibfield
  {title} {\bibinfo {title} {{Perspective: Explicitly correlated electronic
  structure theory for complex systems}},\ }\href
  {https://doi.org/10.1063/1.4976974} {\bibfield  {journal} {\bibinfo
  {journal} {J. Chem. Phys.}\ }\textbf {\bibinfo {volume} {146}},\ \bibinfo
  {pages} {080901} (\bibinfo {year} {2017})}\BibitemShut {NoStop}%
\bibitem [{\citenamefont {Boys}\ \emph
  {et~al.}(1969{\natexlab{a}})\citenamefont {Boys}, \citenamefont {Handy},\
  and\ \citenamefont {Linnett}}]{Boys1969a}%
  \BibitemOpen
  \bibfield  {author} {\bibinfo {author} {\bibfnamefont {S.~F.}\ \bibnamefont
  {Boys}}, \bibinfo {author} {\bibfnamefont {N.~C.}\ \bibnamefont {Handy}},\
  and\ \bibinfo {author} {\bibfnamefont {J.~W.}\ \bibnamefont {Linnett}},\
  }\bibfield  {title} {\bibinfo {title} {A condition to remove the
  indeterminacy in interelectronic correlation functions},\ }\href
  {https://doi.org/10.1098/rspa.1969.0038} {\bibfield  {journal} {\bibinfo
  {journal} {Proc. Roy. Soc. London Ser. A.}\ }\textbf {\bibinfo {volume}
  {309}},\ \bibinfo {pages} {209} (\bibinfo {year}
  {1969}{\natexlab{a}})}\BibitemShut {NoStop}%
\bibitem [{\citenamefont {Boys}\ \emph
  {et~al.}(1969{\natexlab{b}})\citenamefont {Boys}, \citenamefont {Handy},\
  and\ \citenamefont {Linnett}}]{Boys1969b}%
  \BibitemOpen
  \bibfield  {author} {\bibinfo {author} {\bibfnamefont {S.~F.}\ \bibnamefont
  {Boys}}, \bibinfo {author} {\bibfnamefont {N.~C.}\ \bibnamefont {Handy}},\
  and\ \bibinfo {author} {\bibfnamefont {J.~W.}\ \bibnamefont {Linnett}},\
  }\bibfield  {title} {\bibinfo {title} {The determination of energies and
  wavefunctions with full electronic correlation},\ }\href
  {https://doi.org/10.1098/rspa.1969.0061} {\bibfield  {journal} {\bibinfo
  {journal} {Proc. Roy. Soc. London Ser. A.}\ }\textbf {\bibinfo {volume}
  {310}},\ \bibinfo {pages} {43} (\bibinfo {year}
  {1969}{\natexlab{b}})}\BibitemShut {NoStop}%
\bibitem [{\citenamefont {Liao}\ \emph {et~al.}(2021)\citenamefont {Liao},
  \citenamefont {Schraivogel}, \citenamefont {Luo}, \citenamefont {Kats},\ and\
  \citenamefont {Alavi}}]{Liao_transcorrelated}%
  \BibitemOpen
  \bibfield  {author} {\bibinfo {author} {\bibfnamefont {K.}~\bibnamefont
  {Liao}}, \bibinfo {author} {\bibfnamefont {T.}~\bibnamefont {Schraivogel}},
  \bibinfo {author} {\bibfnamefont {H.}~\bibnamefont {Luo}}, \bibinfo {author}
  {\bibfnamefont {D.}~\bibnamefont {Kats}},\ and\ \bibinfo {author}
  {\bibfnamefont {A.}~\bibnamefont {Alavi}},\ }\bibfield  {title} {\bibinfo
  {title} {Towards efficient and accurate ab initio solutions to periodic
  systems via transcorrelation and coupled cluster theory},\ }\href
  {https://doi.org/10.1103/PhysRevResearch.3.033072} {\bibfield  {journal}
  {\bibinfo  {journal} {Phys. Rev. Res.}\ }\textbf {\bibinfo {volume} {3}},\
  \bibinfo {pages} {033072} (\bibinfo {year} {2021})}\BibitemShut {NoStop}%
\bibitem [{\citenamefont {Ten-no}(2000)}]{TENNO2000169}%
  \BibitemOpen
  \bibfield  {author} {\bibinfo {author} {\bibfnamefont {S.}~\bibnamefont
  {Ten-no}},\ }\bibfield  {title} {\bibinfo {title} {A feasible transcorrelated
  method for treating electronic cusps using a frozen gaussian geminal},\
  }\href {https://doi.org/10.1016/S0009-2614(00)01066-6} {\bibfield  {journal}
  {\bibinfo  {journal} {Chem. Phys. Lett.}\ }\textbf {\bibinfo {volume}
  {330}},\ \bibinfo {pages} {169 } (\bibinfo {year} {2000})}\BibitemShut
  {NoStop}%
\bibitem [{\citenamefont {Cohen}\ \emph {et~al.}(2019)\citenamefont {Cohen},
  \citenamefont {Luo}, \citenamefont {Guther}, \citenamefont {Dobrautz},
  \citenamefont {Tew},\ and\ \citenamefont {Alavi}}]{Cohen2019}%
  \BibitemOpen
  \bibfield  {author} {\bibinfo {author} {\bibfnamefont {A.~J.}\ \bibnamefont
  {Cohen}}, \bibinfo {author} {\bibfnamefont {H.}~\bibnamefont {Luo}}, \bibinfo
  {author} {\bibfnamefont {K.}~\bibnamefont {Guther}}, \bibinfo {author}
  {\bibfnamefont {W.}~\bibnamefont {Dobrautz}}, \bibinfo {author}
  {\bibfnamefont {D.~P.}\ \bibnamefont {Tew}},\ and\ \bibinfo {author}
  {\bibfnamefont {A.}~\bibnamefont {Alavi}},\ }\bibfield  {title} {\bibinfo
  {title} {{Similarity transformation of the electronic Schrödinger equation
  via Jastrow factorization}},\ }\href {https://doi.org/10.1063/1.5116024}
  {\bibfield  {journal} {\bibinfo  {journal} {J. Chem. Phys.}\ }\textbf
  {\bibinfo {volume} {151}},\ \bibinfo {pages} {061101} (\bibinfo {year}
  {2019})}\BibitemShut {NoStop}%
\bibitem [{\citenamefont {Feller}(2013)}]{Feller2013}%
  \BibitemOpen
  \bibfield  {author} {\bibinfo {author} {\bibfnamefont {D.}~\bibnamefont
  {Feller}},\ }\bibfield  {title} {\bibinfo {title} {Benchmarks of improved
  complete basis set extrapolation schemes designed for standard ccsd(t)
  atomization energies},\ }\href {https://doi.org/10.1063/1.4791560} {\bibfield
   {journal} {\bibinfo  {journal} {J. Chem. Phys.}\ }\textbf {\bibinfo {volume}
  {138}},\ \bibinfo {pages} {074103} (\bibinfo {year} {2013})}\BibitemShut
  {NoStop}%
\bibitem [{\citenamefont {Irmler}\ \emph {et~al.}(2021)\citenamefont {Irmler},
  \citenamefont {Gallo},\ and\ \citenamefont {Grüneis}}]{Irmler2021}%
  \BibitemOpen
  \bibfield  {author} {\bibinfo {author} {\bibfnamefont {A.}~\bibnamefont
  {Irmler}}, \bibinfo {author} {\bibfnamefont {A.}~\bibnamefont {Gallo}},\ and\
  \bibinfo {author} {\bibfnamefont {A.}~\bibnamefont {Grüneis}},\ }\bibfield
  {title} {\bibinfo {title} {Focal-point approach with pair-specific cusp
  correction for coupled-cluster theory},\ }\href
  {https://doi.org/10.1063/5.0050054} {\bibfield  {journal} {\bibinfo
  {journal} {J. Chem. Phys.}\ }\textbf {\bibinfo {volume} {154}},\ \bibinfo
  {pages} {234103} (\bibinfo {year} {2021})}\BibitemShut {NoStop}%
\bibitem [{\citenamefont {Irmler}\ \emph {et~al.}(2019)\citenamefont {Irmler},
  \citenamefont {Gallo}, \citenamefont {Hummel},\ and\ \citenamefont
  {Gr\"uneis}}]{Irmler2019}%
  \BibitemOpen
  \bibfield  {author} {\bibinfo {author} {\bibfnamefont {A.}~\bibnamefont
  {Irmler}}, \bibinfo {author} {\bibfnamefont {A.}~\bibnamefont {Gallo}},
  \bibinfo {author} {\bibfnamefont {F.}~\bibnamefont {Hummel}},\ and\ \bibinfo
  {author} {\bibfnamefont {A.}~\bibnamefont {Gr\"uneis}},\ }\bibfield  {title}
  {\bibinfo {title} {Duality of ring and ladder diagrams and its importance for
  many-electron perturbation theories},\ }\href
  {https://doi.org/10.1103/PhysRevLett.123.156401} {\bibfield  {journal}
  {\bibinfo  {journal} {Phys. Rev. Lett.}\ }\textbf {\bibinfo {volume} {123}},\
  \bibinfo {pages} {156401} (\bibinfo {year} {2019})}\BibitemShut {NoStop}%
\bibitem [{\citenamefont {Pines}\ and\ \citenamefont
  {Bohm}(1952)}]{pines_collective_1952}%
  \BibitemOpen
  \bibfield  {author} {\bibinfo {author} {\bibfnamefont {D.}~\bibnamefont
  {Pines}}\ and\ \bibinfo {author} {\bibfnamefont {D.}~\bibnamefont {Bohm}},\
  }\bibfield  {title} {\bibinfo {title} {A {Collective} {Description} of
  {Electron} {Interactions}: {II}. {Collective} vs {Individual} {Particle}
  {Aspects} of the {Interactions}},\ }\href
  {https://doi.org/10.1103/PhysRev.85.338} {\bibfield  {journal} {\bibinfo
  {journal} {Phys. Rev.}\ }\textbf {\bibinfo {volume} {85}},\ \bibinfo {pages}
  {338} (\bibinfo {year} {1952})}\BibitemShut {NoStop}%
\bibitem [{\citenamefont {Hede}\ and\ \citenamefont
  {Carbotte}(1972)}]{Hede1972}%
  \BibitemOpen
  \bibfield  {author} {\bibinfo {author} {\bibfnamefont {B.~B.~J.}\
  \bibnamefont {Hede}}\ and\ \bibinfo {author} {\bibfnamefont {J.~P.}\
  \bibnamefont {Carbotte}},\ }\bibfield  {title} {\bibinfo {title}
  {Spin-up-spin-down pair distribution function at metallic densities},\ }\href
  {https://doi.org/10.1139/p72-237} {\bibfield  {journal} {\bibinfo  {journal}
  {Can. J. Phys.}\ }\textbf {\bibinfo {volume} {50}},\ \bibinfo {pages} {1756}
  (\bibinfo {year} {1972})}\BibitemShut {NoStop}%
\bibitem [{\citenamefont {Yasuhara}(1972)}]{yasuhara_1972}%
  \BibitemOpen
  \bibfield  {author} {\bibinfo {author} {\bibfnamefont {H.}~\bibnamefont
  {Yasuhara}},\ }\bibfield  {title} {\bibinfo {title} {Short-range correlation
  in electron gas},\ }\href {https://doi.org/10.1016/0038-1098(72)90504-2}
  {\bibfield  {journal} {\bibinfo  {journal} {Solid State Commun.}\ }\textbf
  {\bibinfo {volume} {11}},\ \bibinfo {pages} {1481} (\bibinfo {year}
  {1972})}\BibitemShut {NoStop}%
\bibitem [{\citenamefont {Singwi}\ \emph {et~al.}(1968)\citenamefont {Singwi},
  \citenamefont {Tosi}, \citenamefont {Land},\ and\ \citenamefont
  {Sj\"olander}}]{Singwi1968}%
  \BibitemOpen
  \bibfield  {author} {\bibinfo {author} {\bibfnamefont {K.~S.}\ \bibnamefont
  {Singwi}}, \bibinfo {author} {\bibfnamefont {M.~P.}\ \bibnamefont {Tosi}},
  \bibinfo {author} {\bibfnamefont {R.~H.}\ \bibnamefont {Land}},\ and\
  \bibinfo {author} {\bibfnamefont {A.}~\bibnamefont {Sj\"olander}},\
  }\bibfield  {title} {\bibinfo {title} {Electron correlations at metallic
  densities},\ }\href {https://doi.org/10.1103/PhysRev.176.589} {\bibfield
  {journal} {\bibinfo  {journal} {Phys. Rev.}\ }\textbf {\bibinfo {volume}
  {176}},\ \bibinfo {pages} {589} (\bibinfo {year} {1968})}\BibitemShut
  {NoStop}%
\bibitem [{\citenamefont {Gell-Mann}\ and\ \citenamefont
  {Brueckner}(1957)}]{gell-mann_correlation_1957}%
  \BibitemOpen
  \bibfield  {author} {\bibinfo {author} {\bibfnamefont {M.}~\bibnamefont
  {Gell-Mann}}\ and\ \bibinfo {author} {\bibfnamefont {K.~A.}\ \bibnamefont
  {Brueckner}},\ }\bibfield  {title} {\bibinfo {title} {Correlation {Energy} of
  an {Electron} {Gas} at {High} {Density}},\ }\href
  {https://doi.org/10.1103/PhysRev.106.364} {\bibfield  {journal} {\bibinfo
  {journal} {Phys. Rev.}\ }\textbf {\bibinfo {volume} {106}},\ \bibinfo {pages}
  {364} (\bibinfo {year} {1957})}\BibitemShut {NoStop}%
\bibitem [{\citenamefont {Qian}(2006)}]{Qian2006}%
  \BibitemOpen
  \bibfield  {author} {\bibinfo {author} {\bibfnamefont {Z.}~\bibnamefont
  {Qian}},\ }\bibfield  {title} {\bibinfo {title} {On-top pair-correlation
  function in the homogeneous electron liquid},\ }\href
  {https://doi.org/10.1103/PhysRevB.73.035106} {\bibfield  {journal} {\bibinfo
  {journal} {Phys. Rev. B}\ }\textbf {\bibinfo {volume} {73}},\ \bibinfo
  {pages} {035106} (\bibinfo {year} {2006})}\BibitemShut {NoStop}%
\bibitem [{\citenamefont {Cioslowski}\ and\ \citenamefont
  {Ziesche}(2005)}]{cioslowski_2005}%
  \BibitemOpen
  \bibfield  {author} {\bibinfo {author} {\bibfnamefont {J.}~\bibnamefont
  {Cioslowski}}\ and\ \bibinfo {author} {\bibfnamefont {P.}~\bibnamefont
  {Ziesche}},\ }\bibfield  {title} {\bibinfo {title} {Applicability of the
  ladder theory to the three-dimensional homogeneous electron gas},\ }\href
  {https://doi.org/10.1103/PhysRevB.71.125105} {\bibfield  {journal} {\bibinfo
  {journal} {Phys. Rev. B}\ }\textbf {\bibinfo {volume} {71}},\ \bibinfo
  {pages} {125105} (\bibinfo {year} {2005})}\BibitemShut {NoStop}%
\bibitem [{\citenamefont {Martin}(2004)}]{Martin2004}%
  \BibitemOpen
  \bibfield  {author} {\bibinfo {author} {\bibfnamefont {R.}~\bibnamefont
  {Martin}},\ }\href {https://doi.org/10.1017/CBO9780511805769} {\emph
  {\bibinfo {title} {Electronic Structure: Basic Theory and Practical
  Methods}}}\ (\bibinfo  {publisher} {Cambridge University Press},\ \bibinfo
  {year} {2004})\BibitemShut {NoStop}%
\bibitem [{\citenamefont {Sundararaman}\ and\ \citenamefont
  {Arias}(2013)}]{Sundararaman2013}%
  \BibitemOpen
  \bibfield  {author} {\bibinfo {author} {\bibfnamefont {R.}~\bibnamefont
  {Sundararaman}}\ and\ \bibinfo {author} {\bibfnamefont {T.~A.}\ \bibnamefont
  {Arias}},\ }\bibfield  {title} {\bibinfo {title} {Regularization of the
  coulomb singularity in exact exchange by wigner-seitz truncated interactions:
  Towards chemical accuracy in nontrivial systems},\ }\href
  {https://doi.org/10.1103/PhysRevB.87.165122} {\bibfield  {journal} {\bibinfo
  {journal} {Phys. Rev. B}\ }\textbf {\bibinfo {volume} {87}},\ \bibinfo
  {pages} {165122} (\bibinfo {year} {2013})}\BibitemShut {NoStop}%
\bibitem [{\citenamefont {Fraser}\ \emph {et~al.}(1996)\citenamefont {Fraser},
  \citenamefont {Foulkes}, \citenamefont {Rajagopal}, \citenamefont {Needs},
  \citenamefont {Kenny},\ and\ \citenamefont {Williamson}}]{Fraser1996}%
  \BibitemOpen
  \bibfield  {author} {\bibinfo {author} {\bibfnamefont {L.~M.}\ \bibnamefont
  {Fraser}}, \bibinfo {author} {\bibfnamefont {W.~M.~C.}\ \bibnamefont
  {Foulkes}}, \bibinfo {author} {\bibfnamefont {G.}~\bibnamefont {Rajagopal}},
  \bibinfo {author} {\bibfnamefont {R.~J.}\ \bibnamefont {Needs}}, \bibinfo
  {author} {\bibfnamefont {S.~D.}\ \bibnamefont {Kenny}},\ and\ \bibinfo
  {author} {\bibfnamefont {A.~J.}\ \bibnamefont {Williamson}},\ }\bibfield
  {title} {\bibinfo {title} {Finite-size effects and coulomb interactions in
  quantum monte carlo calculations for homogeneous systems with periodic
  boundary conditions},\ }\href {https://doi.org/10.1103/PhysRevB.53.1814}
  {\bibfield  {journal} {\bibinfo  {journal} {Phys. Rev. B}\ }\textbf {\bibinfo
  {volume} {53}},\ \bibinfo {pages} {1814} (\bibinfo {year}
  {1996})}\BibitemShut {NoStop}%
\bibitem [{\citenamefont {Shepherd}\ \emph {et~al.}(2014)\citenamefont
  {Shepherd}, \citenamefont {Henderson},\ and\ \citenamefont
  {Scuseria}}]{Shepherd2014}%
  \BibitemOpen
  \bibfield  {author} {\bibinfo {author} {\bibfnamefont {J.~J.}\ \bibnamefont
  {Shepherd}}, \bibinfo {author} {\bibfnamefont {T.~M.}\ \bibnamefont
  {Henderson}},\ and\ \bibinfo {author} {\bibfnamefont {G.~E.}\ \bibnamefont
  {Scuseria}},\ }\bibfield  {title} {\bibinfo {title} {Coupled cluster channels
  in the homogeneous electron gas},\ }\href {https://doi.org/10.1063/1.4867783}
  {\bibfield  {journal} {\bibinfo  {journal} {J. Chem. Phys.}\ }\textbf
  {\bibinfo {volume} {140}},\ \bibinfo {pages} {124102} (\bibinfo {year}
  {2014})}\BibitemShut {NoStop}%
\bibitem [{\citenamefont {Lee}\ and\ \citenamefont
  {Bartlett}(1984)}]{Lee1984a}%
  \BibitemOpen
  \bibfield  {author} {\bibinfo {author} {\bibfnamefont {Y.~S.}\ \bibnamefont
  {Lee}}\ and\ \bibinfo {author} {\bibfnamefont {R.~J.}\ \bibnamefont
  {Bartlett}},\ }\bibfield  {title} {\bibinfo {title} {{A study of Be2 with
  many‐body perturbation theory and a coupled‐cluster method including
  triple excitations}},\ }\href {https://doi.org/10.1063/1.447214} {\bibfield
  {journal} {\bibinfo  {journal} {J. Chem. Phys.}\ }\textbf {\bibinfo {volume}
  {80}},\ \bibinfo {pages} {4371} (\bibinfo {year} {1984})}\BibitemShut
  {NoStop}%
\bibitem [{\citenamefont {Lee}\ \emph {et~al.}(1984)\citenamefont {Lee},
  \citenamefont {Kucharski},\ and\ \citenamefont {Bartlett}}]{Lee1984b}%
  \BibitemOpen
  \bibfield  {author} {\bibinfo {author} {\bibfnamefont {Y.~S.}\ \bibnamefont
  {Lee}}, \bibinfo {author} {\bibfnamefont {S.~A.}\ \bibnamefont {Kucharski}},\
  and\ \bibinfo {author} {\bibfnamefont {R.~J.}\ \bibnamefont {Bartlett}},\
  }\bibfield  {title} {\bibinfo {title} {{A coupled cluster approach with
  triple excitations}},\ }\href {https://doi.org/10.1063/1.447591} {\bibfield
  {journal} {\bibinfo  {journal} {J. Chem. Phys.}\ }\textbf {\bibinfo {volume}
  {81}},\ \bibinfo {pages} {5906} (\bibinfo {year} {1984})}\BibitemShut
  {NoStop}%
\bibitem [{\citenamefont {Urban}\ \emph {et~al.}(1985)\citenamefont {Urban},
  \citenamefont {Noga}, \citenamefont {Cole},\ and\ \citenamefont
  {Bartlett}}]{Urban1985}%
  \BibitemOpen
  \bibfield  {author} {\bibinfo {author} {\bibfnamefont {M.}~\bibnamefont
  {Urban}}, \bibinfo {author} {\bibfnamefont {J.}~\bibnamefont {Noga}},
  \bibinfo {author} {\bibfnamefont {S.~J.}\ \bibnamefont {Cole}},\ and\
  \bibinfo {author} {\bibfnamefont {R.~J.}\ \bibnamefont {Bartlett}},\
  }\bibfield  {title} {\bibinfo {title} {{Towards a full CCSDT model for
  electron correlation}},\ }\href {https://doi.org/10.1063/1.449067} {\bibfield
   {journal} {\bibinfo  {journal} {J. Chem. Phys.}\ }\textbf {\bibinfo {volume}
  {83}},\ \bibinfo {pages} {4041} (\bibinfo {year} {1985})}\BibitemShut
  {NoStop}%
\bibitem [{\citenamefont {Noga}\ \emph {et~al.}(1987)\citenamefont {Noga},
  \citenamefont {Bartlett},\ and\ \citenamefont {Urban}}]{Noga1987b}%
  \BibitemOpen
  \bibfield  {author} {\bibinfo {author} {\bibfnamefont {J.}~\bibnamefont
  {Noga}}, \bibinfo {author} {\bibfnamefont {R.~J.}\ \bibnamefont {Bartlett}},\
  and\ \bibinfo {author} {\bibfnamefont {M.}~\bibnamefont {Urban}},\ }\bibfield
   {title} {\bibinfo {title} {Towards a full ccsdt model for electron
  correlation. ccsdt-n models},\ }\href
  {https://doi.org/https://doi.org/10.1016/0009-2614(87)87107-5} {\bibfield
  {journal} {\bibinfo  {journal} {Chem. Phys. Lett.}\ }\textbf {\bibinfo
  {volume} {134}},\ \bibinfo {pages} {126} (\bibinfo {year}
  {1987})}\BibitemShut {NoStop}%
\bibitem [{\citenamefont {Raghavachari}\ \emph {et~al.}(1989)\citenamefont
  {Raghavachari}, \citenamefont {Trucks}, \citenamefont {Pople},\ and\
  \citenamefont {Head-Gordon}}]{Raghavachari1989}%
  \BibitemOpen
  \bibfield  {author} {\bibinfo {author} {\bibfnamefont {K.}~\bibnamefont
  {Raghavachari}}, \bibinfo {author} {\bibfnamefont {G.~W.}\ \bibnamefont
  {Trucks}}, \bibinfo {author} {\bibfnamefont {J.~A.}\ \bibnamefont {Pople}},\
  and\ \bibinfo {author} {\bibfnamefont {M.}~\bibnamefont {Head-Gordon}},\
  }\bibfield  {title} {\bibinfo {title} {A fifth-order perturbation comparison
  of electron correlation theories},\ }\href
  {https://doi.org/https://doi.org/10.1016/S0009-2614(89)87395-6} {\bibfield
  {journal} {\bibinfo  {journal} {Chem. Phys. Lett.}\ }\textbf {\bibinfo
  {volume} {157}},\ \bibinfo {pages} {479} (\bibinfo {year}
  {1989})}\BibitemShut {NoStop}%
\bibitem [{cc4()}]{cc4ueg}%
  \BibitemOpen
  \href {https://github.com/nmasios/cc4ueg.git} {\bibinfo {title} {cc4ueg;
  available on {\tt https://github.com/nmasios/cc4ueg.git}.}}\BibitemShut
  {Stop}%
\bibitem [{\citenamefont {Irmler}\ and\ \citenamefont
  {Grüneis}(2019)}]{Irmler2019a}%
  \BibitemOpen
  \bibfield  {author} {\bibinfo {author} {\bibfnamefont {A.}~\bibnamefont
  {Irmler}}\ and\ \bibinfo {author} {\bibfnamefont {A.}~\bibnamefont
  {Grüneis}},\ }\bibfield  {title} {\bibinfo {title} {{Particle-particle
  ladder based basis-set corrections applied to atoms and molecules using
  coupled-cluster theory}},\ }\href {https://doi.org/10.1063/1.5110885}
  {\bibfield  {journal} {\bibinfo  {journal} {J. Chem. Phys.}\ }\textbf
  {\bibinfo {volume} {151}},\ \bibinfo {pages} {104107} (\bibinfo {year}
  {2019})}\BibitemShut {NoStop}%
\bibitem [{\citenamefont {Mattuck}(1992)}]{mattuck}%
  \BibitemOpen
  \bibfield  {author} {\bibinfo {author} {\bibfnamefont {R.~D.}\ \bibnamefont
  {Mattuck}},\ }\href@noop {} {\emph {\bibinfo {title} {A guide to {Feynman}
  diagrams in the many-body problem}}},\ \bibinfo {edition} {2nd}\ ed.,\ Dover
  books on physics and chemistry\ (\bibinfo  {publisher} {Dover Publications},\
  \bibinfo {address} {New York},\ \bibinfo {year} {1992})\BibitemShut {NoStop}%
\bibitem [{\citenamefont {Martin}\ \emph {et~al.}(2016)\citenamefont {Martin},
  \citenamefont {Reining},\ and\ \citenamefont {Ceperley}}]{Martin2016}%
  \BibitemOpen
  \bibfield  {author} {\bibinfo {author} {\bibfnamefont {R.}~\bibnamefont
  {Martin}}, \bibinfo {author} {\bibfnamefont {L.}~\bibnamefont {Reining}},\
  and\ \bibinfo {author} {\bibfnamefont {D.}~\bibnamefont {Ceperley}},\ }\href
  {https://books.google.at/books?id=ch1CDAAAQBAJ} {\emph {\bibinfo {title}
  {Interacting Electrons}}}\ (\bibinfo  {publisher} {Cambridge University
  Press},\ \bibinfo {year} {2016})\BibitemShut {NoStop}%
\bibitem [{\citenamefont {Liao}\ and\ \citenamefont
  {Gr\"{u}neis}(2016)}]{Liao2016}%
  \BibitemOpen
  \bibfield  {author} {\bibinfo {author} {\bibfnamefont {K.}~\bibnamefont
  {Liao}}\ and\ \bibinfo {author} {\bibfnamefont {A.}~\bibnamefont
  {Gr\"{u}neis}},\ }\bibfield  {title} {\bibinfo {title} {Communication: Finite
  size correction in periodic coupled cluster theory calculations of solids},\
  }\href {https://doi.org/10.1063/1.4964307} {\bibfield  {journal} {\bibinfo
  {journal} {J. Chem. Phys.}\ }\textbf {\bibinfo {volume} {145}},\ \bibinfo
  {pages} {141102} (\bibinfo {year} {2016})}\BibitemShut {NoStop}%
\bibitem [{\citenamefont {Gruber}\ \emph {et~al.}(2018)\citenamefont {Gruber},
  \citenamefont {Liao}, \citenamefont {Tsatsoulis}, \citenamefont {Hummel},\
  and\ \citenamefont {Gr\"{u}neis}}]{Gruber2018}%
  \BibitemOpen
  \bibfield  {author} {\bibinfo {author} {\bibfnamefont {T.}~\bibnamefont
  {Gruber}}, \bibinfo {author} {\bibfnamefont {K.}~\bibnamefont {Liao}},
  \bibinfo {author} {\bibfnamefont {T.}~\bibnamefont {Tsatsoulis}}, \bibinfo
  {author} {\bibfnamefont {F.}~\bibnamefont {Hummel}},\ and\ \bibinfo {author}
  {\bibfnamefont {A.}~\bibnamefont {Gr\"{u}neis}},\ }\bibfield  {title}
  {\bibinfo {title} {Applying the coupled-cluster ansatz to solids and surfaces
  in the thermodynamic limit},\ }\href
  {https://doi.org/10.1103/PhysRevX.8.021043} {\bibfield  {journal} {\bibinfo
  {journal} {Phys. Rev. X}\ }\textbf {\bibinfo {volume} {8}},\ \bibinfo {pages}
  {021043} (\bibinfo {year} {2018})}\BibitemShut {NoStop}%
\bibitem [{\citenamefont {Mihm}\ \emph {et~al.}(2023)\citenamefont {Mihm},
  \citenamefont {Weiler},\ and\ \citenamefont {Shepherd}}]{Mihm2023}%
  \BibitemOpen
  \bibfield  {author} {\bibinfo {author} {\bibfnamefont {T.~N.}\ \bibnamefont
  {Mihm}}, \bibinfo {author} {\bibfnamefont {L.}~\bibnamefont {Weiler}},\ and\
  \bibinfo {author} {\bibfnamefont {J.~J.}\ \bibnamefont {Shepherd}},\
  }\bibfield  {title} {\bibinfo {title} {How the exchange energy can affect the
  power laws used to extrapolate the coupled cluster correlation energy to the
  thermodynamic limit},\ }\href {https://doi.org/10.1021/acs.jctc.2c00737}
  {\bibfield  {journal} {\bibinfo  {journal} {J. Chem. Theory Comput.}\
  }\textbf {\bibinfo {volume} {19}},\ \bibinfo {pages} {1686} (\bibinfo {year}
  {2023})}\BibitemShut {NoStop}%
\bibitem [{\citenamefont {Barnes}\ \emph {et~al.}(2008)\citenamefont {Barnes},
  \citenamefont {Petersson}, \citenamefont {Feller},\ and\ \citenamefont
  {Peterson}}]{Barnes2008}%
  \BibitemOpen
  \bibfield  {author} {\bibinfo {author} {\bibfnamefont {E.~C.}\ \bibnamefont
  {Barnes}}, \bibinfo {author} {\bibfnamefont {G.~A.}\ \bibnamefont
  {Petersson}}, \bibinfo {author} {\bibfnamefont {D.}~\bibnamefont {Feller}},\
  and\ \bibinfo {author} {\bibfnamefont {K.~A.}\ \bibnamefont {Peterson}},\
  }\bibfield  {title} {\bibinfo {title} {{The CCSD(T) complete basis set limit
  for Ne revisited}},\ }\href {https://doi.org/10.1063/1.3013140} {\bibfield
  {journal} {\bibinfo  {journal} {J. Chem. Phys.}\ }\textbf {\bibinfo {volume}
  {129}},\ \bibinfo {pages} {194115} (\bibinfo {year} {2008})}\BibitemShut
  {NoStop}%
\bibitem [{\citenamefont {Knizia}\ \emph {et~al.}(2009)\citenamefont {Knizia},
  \citenamefont {Adler},\ and\ \citenamefont {Werner}}]{Knizia2009}%
  \BibitemOpen
  \bibfield  {author} {\bibinfo {author} {\bibfnamefont {G.}~\bibnamefont
  {Knizia}}, \bibinfo {author} {\bibfnamefont {T.~B.}\ \bibnamefont {Adler}},\
  and\ \bibinfo {author} {\bibfnamefont {H.-J.}\ \bibnamefont {Werner}},\
  }\bibfield  {title} {\bibinfo {title} {Simplified ccsd(t)-f12 methods: Theory
  and benchmarks},\ }\href {https://doi.org/10.1063/1.3054300} {\bibfield
  {journal} {\bibinfo  {journal} {J. Chem. Phys.}\ }\textbf {\bibinfo {volume}
  {130}},\ \bibinfo {pages} {054104} (\bibinfo {year} {2009})}\BibitemShut
  {NoStop}%
\bibitem [{\citenamefont {Kállay}\ \emph {et~al.}(2021)\citenamefont
  {Kállay}, \citenamefont {Horváth}, \citenamefont {Gyevi-Nagy},\ and\
  \citenamefont {Nagy}}]{Kallay2021}%
  \BibitemOpen
  \bibfield  {author} {\bibinfo {author} {\bibfnamefont {M.}~\bibnamefont
  {Kállay}}, \bibinfo {author} {\bibfnamefont {R.~A.}\ \bibnamefont
  {Horváth}}, \bibinfo {author} {\bibfnamefont {L.}~\bibnamefont
  {Gyevi-Nagy}},\ and\ \bibinfo {author} {\bibfnamefont {P.~R.}\ \bibnamefont
  {Nagy}},\ }\bibfield  {title} {\bibinfo {title} {{Size-consistent explicitly
  correlated triple excitation correction}},\ }\href
  {https://doi.org/10.1063/5.0057426} {\bibfield  {journal} {\bibinfo
  {journal} {J. Chem. Phys.}\ }\textbf {\bibinfo {volume} {155}},\ \bibinfo
  {pages} {034107} (\bibinfo {year} {2021})}\BibitemShut {NoStop}%
\bibitem [{\citenamefont {K{\"{o}}hn}(2009)}]{Koehn2009}%
  \BibitemOpen
  \bibfield  {author} {\bibinfo {author} {\bibfnamefont {A.}~\bibnamefont
  {K{\"{o}}hn}},\ }\bibfield  {title} {\bibinfo {title} {{Explicitly correlated
  connected triple excitations in coupled-cluster theory}},\ }\href
  {https://doi.org/10.1063/1.3116792} {\bibfield  {journal} {\bibinfo
  {journal} {J. Chem. Phys.}\ }\textbf {\bibinfo {volume} {130}},\ \bibinfo
  {pages} {131101} (\bibinfo {year} {2009})}\BibitemShut {NoStop}%
\bibitem [{\citenamefont {K{\"{o}}hn}(2010)}]{Koehn2010}%
  \BibitemOpen
  \bibfield  {author} {\bibinfo {author} {\bibfnamefont {A.}~\bibnamefont
  {K{\"{o}}hn}},\ }\bibfield  {title} {\bibinfo {title} {{Explicitly correlated
  coupled-cluster theory using cusp conditions. II. Treatment of connected
  triple excitations}},\ }\href {https://doi.org/10.1063/1.3496373} {\bibfield
  {journal} {\bibinfo  {journal} {J. Chem. Phys.}\ }\textbf {\bibinfo {volume}
  {133}},\ \bibinfo {pages} {174118} (\bibinfo {year} {2010})}\BibitemShut
  {NoStop}%
\end{thebibliography}%

\end{document}